\documentclass[preprint,12pt,3p]{elsarticle}

\usepackage{graphicx}
\usepackage{amssymb}
\usepackage{caption}
\usepackage{subcaption}
\usepackage{url}
\usepackage{amsmath}
\usepackage{float}

\graphicspath{{./Images/}}

\journal{Bioinspiration \& Biomimetics}

\begin{document}

\begin{frontmatter}

\title{\textit{IB2d}: a Python and MATLAB implementation of the immersed boundary method}


\author[label1]{Nicholas A. Battista\corref{cor1}\fnref{label3}}
\address[label1]{Department of Mathematics, CB 3250, University of North Carolina, Chapel Hill, NC, 27599}

\cortext[cor1]{I am corresponding author}

\ead{nick.battista@unc.edu}
\ead[url]{battista.web.unc.edu}

\author[label1,label2]{W. Christopher Strickland}
\address[label2]{SAMSI, T.W. Alexander Drive, P.O. Box 14006, Research Triangle Park, NC 27709}
\ead{cstrickland@samsi.info}

\author[label1,label3]{Laura A. Miller}
\address[label3]{Department of Biology, CB 3280, University of North Carolina, Chapel Hill, NC, 27599}
\ead{lam9@unc.edu}

\begin{abstract}
The development of fluid-structure interaction (FSI) software involves trade-offs between ease of use, generality, performance, and cost. Typically there are large learning curves when using low-level software to model the interaction of an elastic structure immersed in a uniform density fluid. Many existing codes are not publicly available, and the commercial software that exists usually requires expensive licenses and may not be as robust or allow the necessary flexibility that in house codes can provide. We present an open source immersed boundary software package, \textit{IB2d}, with full implementations in both MATLAB and Python, that is capable of running a vast range of biomechanics models and is accessible to scientists who have experience in high-level programming environments. \textit{IB2d} contains multiple options for constructing material properties of the fiber structure, as well as the advection-diffusion of a chemical gradient, muscle mechanics models, and artificial forcing to drive boundaries with a preferred motion. 
\end{abstract}


\end{frontmatter}


%
%

\section{Introduction}
\label{introduction}

Fully coupled fluid-structure interaction problems (FSI) is a rapidly growing discipline across all the sciences, ranging from engineering to biology \cite{Bathe:2008}. Fully coupled FSI is different from models in which the motion or bending of a structure is prescribed. The action of the aortic valve is a good example of fully coupled FSI, since the motion of the valve is governed by the motion of the fluid, and in turn, the valve alters the underlying blood flow. Note that in a fully coupled simulation, the movement of the valve would not be prescribed. 

The immersed boundary ($IB$) framework was first published in 1972 to study blood flow around valve leaflets of the heart by Charles Peskin \cite{Peskin:1972}. It has been applied to a plethora of biomechanics problems which involve the interaction of a flexible structure immersed in a viscous, incompressible fluid. The method has been successfully applied to study fluid dynamics in a variety of biological settings within the intermediate Reynolds number range, defined here as $0.01<Re<1000$, where 
\begin{equation}
\label{Re} Re = \frac{\rho L V}{\mu}.
\end{equation}
$\mu$ and $\rho$ are the dynamic viscosity and density of the fluid, respectively, and $L$ and $V$ are a characteristic length and velocity of the problem. Some of these applications include cardiovascular dynamics \cite{Peskin:1977,GriffithThesis:2005}, aquatic locomotion \cite{Hieber:2008,Hoover:2015}, insect flight \cite{Miller:2004,Miller:2009,SJones:2015}, muscle-fluid-structure interactions \cite{Tytell:2010,Battista:2015,Hamlet:2015}, and plant biomechanics \cite{Zhu:2011}.

The strength of this method is that it can be used to model fully coupled  fluid-structure interaction problems involving complicated time-dependent geometries using a regular fixed Cartesian discretization of the fluid domain, while the elastic fibers describing the structure are discretized on a Lagrangian mesh. The fluid and elastic fibers constitute a coupled system in which the structure moves at the local fluid velocity and the structure applies a singular force to the fluid.

Beyond fully-coupled fluid-structure interaction models, many scientists have successfully coupled other constitutive equations within the $IB$ framework \cite{Kramer:2008,Fogelson:2008,Tytell:2010,Lee:2010,Strychalski:2012,Du:2014,Baird:2015,Waldrop:2015}. For example, in \cite{Fogelson:2008}, Fogelson and Guy modeled platelets suspended in an incompressible fluid to study blood clotting and included chemical reaction equations modeling the mechanisms for binding-unbinding, platelet stimulus-response, and chemistry on the platelets surfaces. Moreover, in \cite{Tytell:2010}, Tytell \textit{et al.} incorporated calcium dynamics, which governed the muscle contraction dynamics that were responsible for force generation in a swimming lamprey.  

Many implementations of the immersed boundary method ($IBM$) exist in compiled programming languages, including a few open source and freely available packages, e.g., IBIS \cite{Fogelson:IBIS} and IBAMR \cite{BGriffithIBAMR}. IBIS is $IB$ software written in FORTRAN that includes its own graphical user interface, \textit{ibisview}, to visualize the simulations. IBAMR is an adaptive and parallelized implementation of the $IBM$ in $C++$, with extensions to a hybrid implementation of $IB$ which uses a finite element discretization of the immersed structure \cite{Griffith:IBFE}. It depends on many open source libraries, including PETSc \cite{PETSc}, SAMRAI \cite{SAMRAI}, libMesh \cite{libMesh}, and OpenMPI \cite{openMPI} which make it robust and very efficient to run but at the cost of a steep learning curve for anyone inexperienced at high performance computing. Moreover installation of IBAMR is non-trivial, as it requires installing the above open source libraries and coupling them with the IBAMR framework. Furthermore, without multi-processor computational resources available, IBAMR cannot run at its full potential.

IBAMR was developed for highly resolved computational grids and specifically designed to include adaptive mesh refinement (AMR) capabilities. AMR allows for more computational speedup; it dynamically adapts the computational grid for higher resolution near regions of interest, e.g., boundaries and regions of vorticity above a user-prescribed threshold, while solving at lower resolution in other areas. Because of IBAMR's AMR and parallelization capabilities, it can be used for $3D$ applications unlike previous open source $IB$ software, such as IBIS, which was strictly developed for $2D$ applications.

IBIS AND IBAMR, having been written in lower level programming languages, e.g., FORTRAN and $C++$ respectively, require familiarity with these languages. For students and scientists from disciplines that are not typically trained in rigorous programming, these languages are often inaccessible and necessitate a steep learning curve.

Recently there have been a few open source $2D$ $IB$ codes available on GitHub, such as matIB \cite{Wiens:MATIB} and pyIBM \cite{Mesnard:pyIBM}, which are a MATLAB and Python 3.5 implementation, respectively. Charlie Peskin also has a MATLAB $2D$ $IB$ implementation available on his website \cite{Peskin:IB}. All these implementations use the standard immersed boundary framework \cite{Peskin:2002} but do not include a breadth of fiber models or examples and are not as robust or efficient in comparison to their 3D counterparts, such as IBAMR. However, implementations in these high-level programming languages offer many powerful advantages, perhaps foremost being that they are more readable and familiar to a broad audience of scientists and engineers. 

In this paper, we present $IBM$ software called \textit{IB2d} with full implementations in both MATLAB \cite{MATLAB:2015a} and Python 3.5 \cite{Python:Python} that is capable of modeling a broad array of problems in biomechanics including (but not limited to) locomotion, physiological processes, and plant biomechanics. Even for skilled programmers, \textit{IB2d} represents a nice option for preliminary tests of new models. For example, one may add new muscle models to the software quite easily for testing before attempting an implementation in a more challenging software framework such as IBAMR. 

\textit{IB2d} is an extension of the preliminary code found in \cite{Battista:2015}. It extends the capabilities of this code by introducing a full implementation in Python, numerous additions in functionality, such as more fiber-structure modelling options, advection-diffusion, electrophysiology models, and artificial forcing, as well as visualization output and data analysis options. The package also contains $30$+ examples, which illustrate the breadth of the software.

%
%

\section{IBM Framework}
\label{IBM_Framework}

\textit{IB2d} models the fluid motion in two dimensions using the Navier-Stokes equations in Eulerian form, given as

\begin{equation}
\label{Navier_Stokes} \rho \left( \frac{\partial {\bf{u}}({\bf{x}},t) }{\partial t} + {\bf{u}}({\bf{x}},t)\cdot \nabla {\bf{u}}({\bf{x}},t) \right) = -\nabla p({\bf{x}},t) + \mu \Delta {\bf{u}}({\bf{x}},t) + {\bf{f}}({\bf{x}},t) \\
\end{equation}
\begin{equation}
\label{Incompressibility} \nabla\cdot {\bf{u}}({\bf{x}},t) = 0,\\
\end{equation}
where ${\bf{u}}({\bf{x}},t) = (u({\bf{x}},t),v({\bf{x}},t))$ is the fluid velocity, $p({\bf{x}},t)$ is the pressure, and ${\bf{f}}({\bf{x}},t)$ is the force per unit volume (area in $2D$) applied to the fluid by the immersed boundary. The independent variables are the position, ${\bf{x}}= (x,y)$, and time, $t$. Eq.(\ref{Navier_Stokes}) is equivalent to the conservation of momentum for a fluid, while Eq.(\ref{Incompressibility}) is the condition mandating that the fluid is incompressible. \textit{IB2d} also assumes a periodic and square fluid domain. Future implementations will include a projection method solver to incorporate Dirichlet and Neumann boundary conditions \cite{Chorin:1967,Brown:2001}.

The interaction equations between the fluid and the immersed structure are given by
\begin{equation}
\label{IBM_Force} {\bf{f}}({\bf{x}},t) = \int {\bf{F}}(r,t)\delta({\bf{x}}-{\bf{X}}(r,t)) dr
\end{equation}
\begin{equation}
\label{IBM_Velocity} {\bf{U}}({\bf{X}}(r,t),t) = \frac{\partial {\bf{X}}(r,t)}{\partial t} = \int {\bf{u}}({\bf{x}},t) \delta( {\bf{x}} - {\bf{X}}(r,t) ) d{\bf{x}},
\end{equation}
where ${\bf{X}}(r,t)$ gives the Cartesian coordinates at time $t$ of the material point labeled by Lagrangian parameter $r$, ${\bf{f}}(r,t)$ is the force per unit area imposed onto the fluid by elastic deformations in the immersed structure as a function of the Lagrangian position, $r$, and time, $t$. The force density, ${\bf{f}}(r,t)$, is a functional of the current immersed boundary's configuration. Moreover, we write the force density as
\begin{equation}
\label{IBM_Force_Density} {\bf{F}}(r,t) = \bf{\mathbb{F}}{(\bf{X}}(r,t),t),
\end{equation}
where $\bf{\mathbb{F}}({\bf{X}},t)$ is a combination of all the fiber components modeling the desired material properties of the immersed structure. The fiber models implemented in \textit{IB2d} are described in subsequent sections. 

Eq.(\ref{IBM_Force}) applies a force from the immersed boundary to the fluid through a delta-kernel integral transformation. Eq.(\ref{IBM_Velocity}) sets the velocity of the boundary equal to the local fluid velocity, to satisfy the no-slip condition.

Upon discretizing Eqs.(\ref{IBM_Force}) and (\ref{IBM_Velocity}), regularized delta functions, $\delta_h$, are implemented,  
\begin{equation}
\label{IBM:delta} \delta_h(\mathbf{x}) = \frac{1}{h^2} \phi\left( \frac{x}{h} \right) \phi\left( \frac{y}{h} \right),
\end{equation}
where $h$ is the fluid grid width and
\begin{equation}
\label{IBM:delta2} \phi(r) = \left\{\begin{array}{c} \frac{1}{4} \left(1 + \cos\left(\frac{\pi r}{2} \right) \right) \ \ \ \ \ \ |r|\leq 2 \\ 0  \ \ \ \ \ \ \ \ \ \  \ \ \ \ \ \ \ \ \ \ \ \ \ \ \ \ \ \ \ \  \mbox{otherwise} \end{array}\right.,
\end{equation}
where $r$ is the distance from the Lagrangian node. More details on regularized delta functions and the discretization can be found in \cite{Peskin:2002}.

The coupled equations (\ref{Navier_Stokes}-\ref{IBM_Velocity}) are solved using the algorithm described in Peskin's $IB$ review paper \cite{Peskin:2002} with periodic boundary conditions imposed on both the fluid and immersed boundary. Details on the discretization used in \textit{IB2d} are found in \ref{appendix:fluidsolve}.

\begin{figure}
    \begin{subfigure}{0.5\textwidth}
        \centering
        \includegraphics[width=0.8\textwidth]{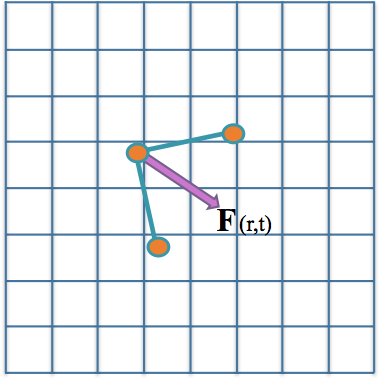}
        \caption{}
        \label{IB_1}
    \end{subfigure}
    \begin{subfigure}{0.5\textwidth}
        \centering
        \includegraphics[width=0.8\textwidth]{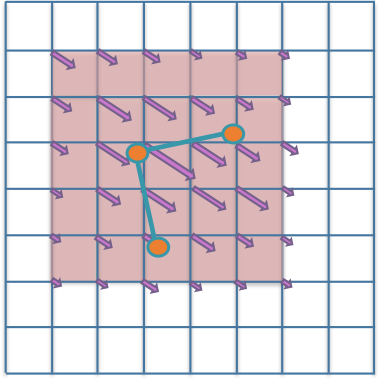}
        \caption{}
        \label{IB_2}
    \end{subfigure}\\ \\
        \begin{subfigure}{0.5\textwidth}
        \centering
        \includegraphics[width=0.8\textwidth]{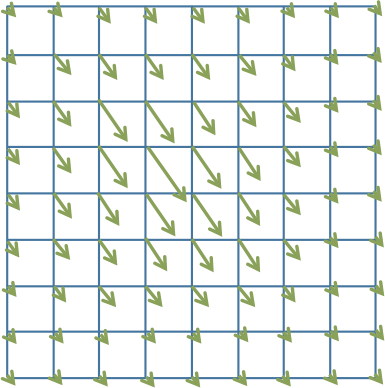}
        \caption{}
        \label{IB_3}
    \end{subfigure}
    \begin{subfigure}{0.5\textwidth}
        \centering
        \includegraphics[width=0.8\textwidth]{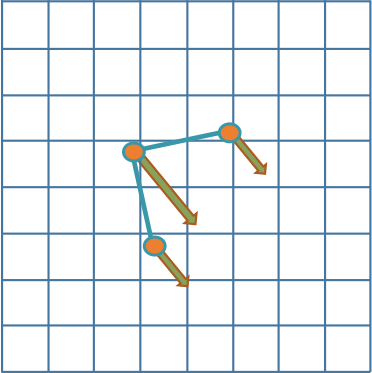}
        \caption{}
        \label{IB_4}
    \end{subfigure}
\caption{A visual guide to the standard steps in Peskin's immersed boundary method. (a) The elastic deformation forces are computed from the current configuration of the immersed structure. (b) Those deformation forces are spread to neighboring fluid grid points, via Eq.(\ref{IBM_Force}). (c) The fluid velocity is updated everywhere in the domain using Eqs.(\ref{Navier_Stokes}) and (\ref{Incompressibility}). (d) The immersed boundary is moved at the local fluid velocity by Eq.(\ref{IBM_Velocity}). Note that the deformation force vectors in (b) and velocity vectors in (c) are not parallel, as the fluid already may have some underlying non-zero velocity field, which gets perturbed due to the presence of the deformation forces.}
\label{IB_Steps}
\end{figure}

The standard numerical algorithm for immersed boundary \cite{Peskin:2002}, illustrated in Figure \ref{IB_Steps}, is as follows:
\begin{itemize}
\item[\emph{Step 1:}] Compute the force density ${\bf{F}}^{n}(r,t)$ on the immersed boundary from the current boundary deformations, ${\bf{X}}^{n}$.
\item[\emph{Step 2:}] Use Eq.(\ref{IBM_Force}) to spread these deformation forces from the Lagrangian nodes to the fluid lattice points nearby.
\item[\emph{Step 3:}] Solve the Navier-Stokes equations, Eqs.(\ref{Navier_Stokes}) and (\ref{Incompressibility}), on the Eulerian domain. E.g., update ${\bf{u}}^{n+1}$ and $p^{n+1}$ from ${\bf{u}}^{n}$ and ${\bf{f}}^{n}$. Note: since we are enforcing periodic boundary conditions on the computational domain, the Fast Fourier Transform (FFT) \cite{Cooley:1965,Press:1992} is used to solve for these updated quantities at an accelerated rate.
\item[\emph{Step 4:}] Update the fiber model positions, ${\bf{X}}^{n+1}$, using the local fluid velocities, ${\bf{U}}^{n+1}$, using ${\bf{u}}^{n+1}$ and Eq.(\ref{IBM_Velocity}). E.g., move the immersed structure at the local fluid velocities thereby enforcing \emph{no slip} boundary conditions.
\end{itemize}

%
%

\subsection{Fiber Models}
\label{FiberModels}

In this section, all current fiber models implemented in \textit{IB2d} are described. Various fiber models give the immersed boundary certain desirable material properties relevant to many scientific applications. Currently the following types of fiber models are available:
\begin{enumerate}
\item Springs (Hookean or Non-Hookean)
\item Torsional Springs
\item Target Points
\item Mass Points (with or without gravity)
\item Porosity
\item Muscle-Fluid-Structure Models
\end{enumerate}

Once the deformation energy has been calculated in the algorithm, e.g., in Step $3$, 
\begin{equation}
\label{ElasticEnergy} E({(\bf{X}}(r,t),t) = \sum_{k=0}^{M} E_k( {\bf{X}}_{k,1}, {\bf{X}}_{k,2}, \ldots {\bf{X}}_{k,N_k}),
\end{equation}
the corresponding elastic forces can be computed via derivatives of the elastic energy, where the elastic deformation force at point $c$ of fiber model $k$ is calculated as
\begin{equation}
\label{ElasticForce} \bf{\mathbb{F}}_{k,c}{(\bf{X}}(r,t),t) = - \frac{\partial E({\bf{X}}(r,t),t) }{\partial {\bf{X}}_{k,c} }.
\end{equation}

Note that ${\bf{X}}$ contains the coordinates of all immersed boundary points, $M$ is the number of fiber structures in the system, $N_k$ is the number of immersed boundary points in fiber structure $M$, and the negative sign is chosen to drive the system towards a minimal energy state. Furthermore we note that (\ref{ElasticEnergy}) is a combination of the deformation energies from all respective fiber models, which are described below.

%
%
%
%
\begin{figure}
    \begin{subfigure}{0.35\textwidth}
        \centering
        \includegraphics[width=0.95\textwidth]{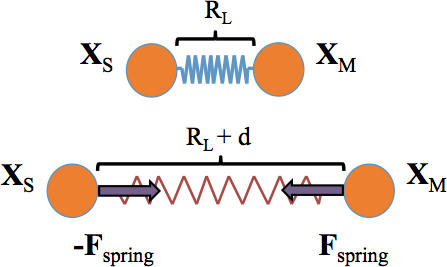}
        \caption{}
        \label{IB_Springs}
    \end{subfigure}
    \begin{subfigure}{0.65\textwidth}
        \centering
        \includegraphics[width=0.95\textwidth]{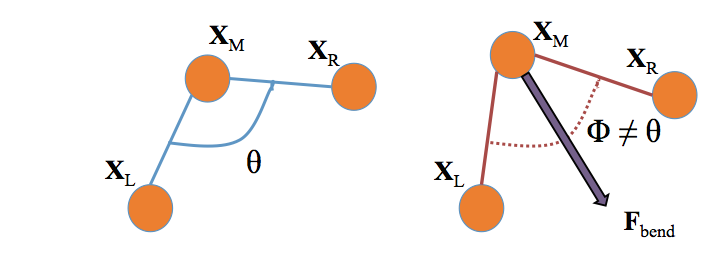}
        \caption{}
        \label{IB_TSprings}
    \end{subfigure}\\ \\
    %
    %
    \begin{subfigure}{\textwidth}
        \centering
        \includegraphics[width=0.95\textwidth]{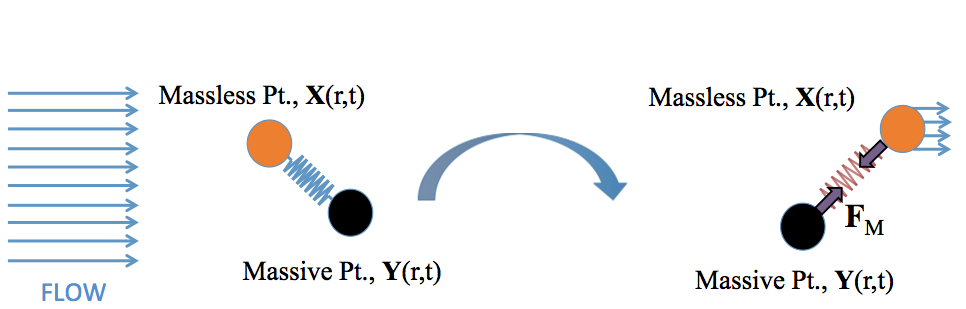}
        \caption{}
        \label{IB_Mass} 
    \end{subfigure} \\ \\
    \begin{subfigure}{\textwidth}
        \centering
        \includegraphics[width=0.45\textwidth]{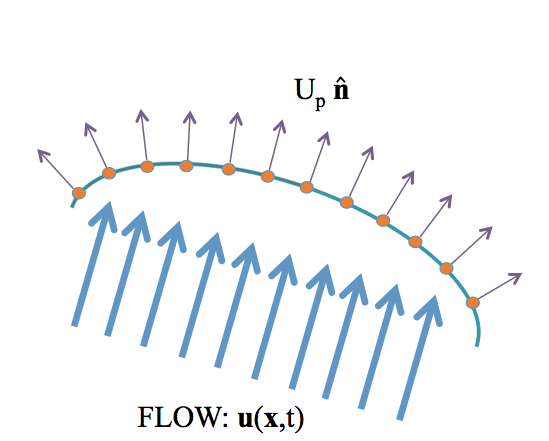}
        \caption{}
        \label{IB_Porous}
    \end{subfigure}
\caption{Illustrating the key points of various fiber models implemented in \textit{IB2D}. (a) Two nodes connected by a virtual spring held at the resting-length of the spring (top) and a rendering of the longitudinal forces induced when the spring is stretched (bottom). (b) A torsional spring connecting three adjacent Lagrangian nodes at its equilibrium configuration, e.g. angle $\theta$ (left) and an illustration of the force experienced by the middle node, $\mathbf{X}_M$, when the system is not at its lowest energy state (right). (c) A massless and massive point, $\mathbf{X}$ and $\mathbf{Y}$ respectively, connected by a stiff virtual spring. Incoming flow moves the massless point to a new position, which exerts a pulling-like effect on the massive point. The massive point will move depending on a coupled constitutive equation. (d) Incoming flow permeates a porous boundary. The amount of flow that moves through the body depends on the permeability of the membrane; all flow through the boundary is normal to the body itself.}
\label{IB_Fiber_Pics}
\end{figure}

%
%
%
%
\subsubsection{Springs}

Resistance to stretching between successive Lagrangian points can be achieved by modeling the connections with Hookean (or Non-Hookean) springs of resting length $R_L$ and spring stiffness $k_S$. If the virtual spring displacement is below or beyond $R_L$, the model will drive the system back towards a lower energy state. The elastic potential energy for a Hookean spring is given by
\begin{equation}
\label{spring:energy} E_{spring} = \frac{1}{2} k_S\ ( ||{\bf{X}}_{SL}-{\bf{X}}_{M}|| - R_L)^{2},
\end{equation}
where ${\bf{X}}_{M}$ and ${\bf{X}}_{SL}$ are master and slave node coordinates respectively. The corresponding deformation force is given by a derivative on the elastic energy as in Eq.(\ref{ElasticForce})
\begin{equation}
\label{spring:force1} F_{spring} = k_S \left( 1 - \frac{R_L}{||{\bf{X}}_{SL}-{\bf{X}}_{M}||} \right) \cdot \left( \begin{array}{c} x_{SL} - x_M \\ y_{SL} - y_M \end{array} \right).
\end{equation}

This fiber model is illustrated in Figure \ref{IB_Springs}. Two nodes are connected by a virtual spring shown when the spring is at resting-length (top). A rendering of the longitudinal forces induced when the spring is stretched is also depicted (bottom).

Furthermore, \textit{IB2d} also implements nonlinear springs that assume the nonlinear extension of Eq.(\ref{spring:energy}), i.e.
\begin{equation}
\label{nlspring:energy} E_{spring} = \frac{1}{2} k_S\ ( ||{\bf{X}}_{SL}-{\bf{X}}_{M}|| - R_L)^{\beta+1},
\end{equation}
where $\beta\in\mathbb{Z}^+$. The corresponding force is then given by
\begin{equation}
\label{spring:force2} F_{spring} = \frac{\beta+1}{2}\ k_S \left( 1 - \frac{R_L}{||{\bf{X}}_{SL}-{\bf{X}}_{M}||} \right)^{\beta}\cdot \left( \begin{array}{c} x_{SL} - x_M \\ y_{SL} - y_M \end{array} \right).
\end{equation}

%
%
%
%
\subsubsection{Torsional Springs (``Beams")}

Resistance to bending between three successive Lagrangian points is modeled using a torsional spring connecting the three nodes. The model assumes a desired angle $\theta$, a prescribed `curvature' between the three Lagrangian points, with corresponding bending stiffness $k_B$. Hence the bending energy is given as
\begin{equation}
\label{tspring:energy} E_{bend} = \frac{1}{2} k_B\ ( \hat{z} \cdot ( \mathbf{X}_R - \mathbf{X}_M ) \times ( \mathbf{X}_M - \mathbf{X}_L )  - C)^{2},
\end{equation}
where $\mathbf{X}_R, \mathbf{X}_L,$ and $\mathbf{X}_M$ are right, left, and master Lagrangian mode coordinates, and $C = d_{LM}d_{MR} \sin\theta.$ Note that $C$ is not the standard definition of curvature, but a curvature defined at the desired angle $\theta$ and distances between links, $d_{LM}$ and $d_{MR}$. 

The penalty force is designed to drive any deviations in the angle between these links back towards a lower energy state, i.e., $\theta$. The corresponding bending force is given by 

{\small
\begin{equation}
\label{tspring:force} \mathbf{F}_{bend} = k_B \Bigg( (x_R -x_M)(y_M-y_L) - (y_R-y_M)(x_M-x_L) - C  \Bigg) \cdot \left( \begin{array}{c} $ $\\ (y_M-y_L)+(y_R-y_M) \\  $ $ \\ -(x_R-x_M) - (x_M-x_L) \\ $ $\\    \end{array} \right)
\end{equation}
}

An illustration of $2D$ torsional springs is shown in Figure \ref{IB_TSprings}, where a torsional spring connects three adjacent Lagrangian nodes $\mathbf{X}_L, \mathbf{X}_M,$ and $\mathbf{X}_R$ at their equilibrium configuration with angle $\theta$ (left). The force is experienced by the middle node, $\mathbf{X}_M$, when the system is not at its lowest energy state (right), driving it back to its preferred configuration.

%
%
%
%
\subsubsection{Target Points}

Target points can be used to prescribe a preferred position or motion of the Lagrangian points. In this formulation, each Lagrangian point is associated with a \emph{virtual} or target point. The boundary point is connected to its virtual target point via a stiff spring, i.e., a spring with zero resting length. Essentially the virtual point mandates where the target point should be. 
The deformation energy is given similarly to Eq.(\ref{spring:energy}),
\begin{equation}
\label{target:energy} E_{T}(\mathbf{X}_M) = \frac{1}{2} k_T \Big|\Big| \mathbf{X}_M - \mathbf{X}_M^T \Big|\Big|^2,
\end{equation}
where $k_T$ is the target point stiffness and $\mathbf{X}_M$ and $\mathbf{X}_M^T$ are the coordinates of the physical Lagrangian point and virtual target point, respectively. Hence the corresponding deformation forces are given as
\begin{equation}
\label{target:force} \mathbf{F}_T = -\frac{\partial E_T}{\partial \mathbf{x}_M} = -k_T \ \left( \begin{array}{c}  x_M-x_M^T \\ y_M - y_M^T  \end{array} \right).
\end{equation}

Note that in both cases, it is standard for $k_T$ to be very large in order to hold the Lagrangian points nearly rigid or move them in a prescribed manner based on updating the positions of the virtual nodes. Many scientists have used this formulation to prescribe motion in a variety of contexts \cite{Miller:2009,Hamlet:2012}.

%
%
%
%
\subsubsection{Massive Points}

Artificial mass can be modeled on the fiber structure using an approach that is similar to target points. $\mathbf{Y}(r,t)$ gives the Cartesian coordinates of the \emph{massive} points, with associated mass density $M(r)$. These points do not interact with the fluid directly and can be thought to be a virtual point. $\mathbf{X}(r,t)$ give the Cartesian coordinates of the Lagrangian boundary points which are massless and interact with the fluid. Recall that the boundary points also move at the local fluid velocity, and exert elastic deformation forces to the local fluid grid. If the massive points deviate from the Lagrangian boundary points, a restoring force will drive them back together. 

The equations modeling this system are
\begin{align}
\label{mass:force} \mathbf{F}_{M} &= k_M ( \mathbf{Y}(r,t) - \mathbf{X}(r,t) ) \\
\label{mass:gravity} M(r) \frac{\partial^2 \mathbf{Y}(r,t) }{\partial t^2} &= -\mathbf{F}_M - M(r) g \hat{e}_2,
\end{align}
where $k_M$ is a stiffness coefficient with $k_M>>1$, and $g$ is the acceleration due to gravity in direction $\hat{e}_2$.

Note that the coupling is very similar to the target point formulation with the distinct difference that, rather than the movement of the massive points being prescribed, it is based on a constitutive equation, Eq.(\ref{mass:gravity}). Furthermore, gravity does not have to be applied in Eq.(\ref{mass:gravity}); rather, the system can be modeled by purely artificial mass alone without the influence of gravity.

A simple rendering of the massive point fiber model is depicted in Figure \ref{IB_Mass}. The Lagrangian boundary point and massive point, with Cartesian coordinates $\mathbf{X}$ and $\mathbf{Y}$, are shown respectively, connected by a stiff virtual spring. Background fluid flow potential moves the massless point to a new position which exerts a pulling-like effect on the massive point. The massive point will move depending on the coupled constitutive equation given in Eq.(\ref{mass:gravity}).

%
%
%
%
\subsubsection{Porosity}

An interpretation of Darcy's Law is used to make the immersed structure permeable to fluid. In other words, the porous structure allows fluid to flow through it. Darcy's Law is a phenomenologically derived constitutive equation, which states the velocity of the fluid flowing through a porous medium is proportional to a pressure gradient of the two sides of the medium. This relation can be written as
\begin{equation}
\label{porosity:darcy1} U_p = -\frac{\kappa_p [p]}{\mu a},
\end{equation}
where $U_p \hat{n}$ is the porous slip velocity and $\kappa_p$ is the membrane permeability, $\mu$ is the fluid's dynamic viscosity, $a$ is the structure's thickness, $[p]$ is the pressure gradient across the boundary, and $\hat{n}$ is a unit vector normal to the structure. However, the pressure jump may be simplified by first integrating across Eq.(\ref{Navier_Stokes}) to eliminate the singular forcing term and obtain jump conditions for the normal and tangential fluid stresses across the boundary, which can be simplified to reduce the pressure jump to
\begin{equation}
\label{porosity:pressure} [p] = \frac{\mathbf{F}\cdot\hat{n}}{|\mathbf{X}_r|},
\end{equation}
as in \cite{Stockie:2009},\cite{Peskin:1993}. Hence the porous slip velocity is found to be
\begin{equation}
\label{porous:darcy} U_p = -\frac{\alpha \mathbf{F}\cdot\hat{n} }{|\mathbf{X}_r|},
\end{equation}
where $\alpha=\frac{\kappa_p}{\mu a}$ is the porous slip parameter and $\mathbf{X}_r$ is the position of the porous Lagrangian structure. As stated in \cite{Stockie:2009}, since $\kappa$ can be easily obtained from experiments, $\alpha$ can be easily found as well. 

Once the Darcy porous slip velocity, e.g. Eq.(\ref{porous:darcy}), is found, Eq.(\ref{IBM_Velocity}) must be adjusted to account for the porosity
\begin{equation}
\label{porous:IBM_Velocity} {\bf{U}}({\bf{X}}(r,t),t) = -U_p \hat{n} +  \int {\bf{u}}({\bf{x}},t) \delta( {\bf{x}} - {\bf{X}}(r,t) ) d{\bf{x}}.
\end{equation}
This formulation was first described in \cite{Kim:2006} and the discretization used to find the normal vectors in \textit{IB2d} can be found in \ref{appendix:porous}. An illustration of porosity is shown in Figure \ref{IB_Porous}, where incoming fluid flow permeates a porous boundary. The amount of flow that moves through the body depends on the permeability parameter, $\alpha$, associated with the membrane. All flow through the boundary is normal to the body itself.

%
%
%
%
\subsubsection{Muscle-Fluid-Structure Models: FV-LT Model}

The simple muscle model described in \cite{Battista:2015} has been incorporated into \textit{IB2d}. This muscle model attempts to model both a force-velocity and length-tension relationship in muscle without coupling in the underlying cellular processes like calcium signaling, myosin cross-bridge attachment and detachment, or filament compliance. 

\begin{figure}
    \begin{subfigure}{0.4\textwidth}
        \centering
        \includegraphics[width=0.95\textwidth]{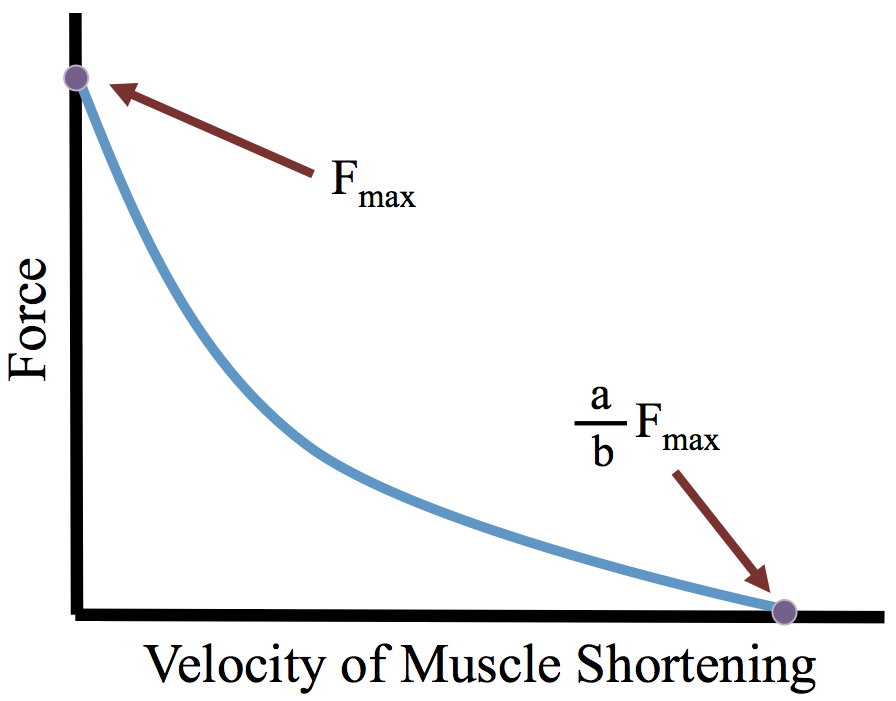}
        \caption{}
        \label{IB_FV_Curve} 
    \end{subfigure} 
    \begin{subfigure}{0.6\textwidth}
        \centering
        \includegraphics[width=0.925\textwidth]{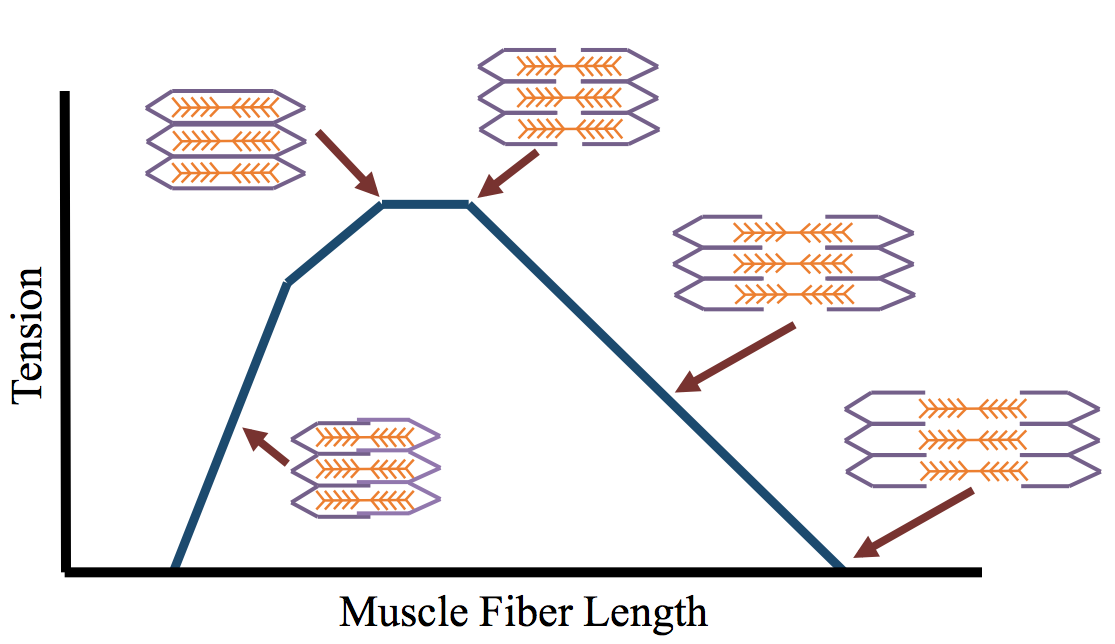}
        \caption{}
        \label{IB_LT_Curve}
    \end{subfigure}
\caption{Example force-velocity and length-tension curves illustrating the respective relationships that the $FV$-$LT$ model is trying to capture.}
\label{FV_LT_Pics}
\end{figure}

The force a muscle can generate depends on the speed of muscle contraction; e.g., the faster a muscle shortens, the less force it generates. Traditionally a Hill model is used to describe this relationship and takes the following form \cite{Hill:1938,Fung:1993},
\begin{equation}
\label{muscle:hill} V_F = \frac{b(F_{max}-F)}{F+a},
\end{equation}
where $V_F$ is the muscle fiber's shortening velocity, $F$ is the force generated by the fiber, and $F_{max}$ is the maximum load at zero contractile velocity. Parameters $a$ and $b$ can be determined experimentally and are are related to the internal thermodynamics of the muscle. An example force-velocity curve is shown in Figure \ref{IB_FV_Curve}.

The force a muscle fiber can generate is also known to be a function of its length. Initially when the thick filaments begin to bind to the thin filaments, the resulting force increases as the muscle shortens. However, if the muscle is contracted too far, there are fewer myosin heads to attach to the actin filaments, and the resulting force exerted is smaller. Hence the maximal muscle tension is generated between the two extremes, i.e., when the myosin heads are within reach of the thin filaments. An example length-tension curve is shown in Figure \ref{IB_LT_Curve} where actin and myosin binding is depicted at varying muscle lengths. A simple model of a length-tension relationship is described in \cite{Hatze:1981}, 
\begin{equation}
\label{muscle:LT} F_I = F_{IO} \exp\left[-\left( \frac{Q-1}{SK} \right)^2 \right],
\end{equation}
where $Q = \frac{L_F}{L_{FO}}$ is the ratio of the length of the muscle fibers to their length when they generate their maximum tension, $F_I$ is the maximum isometric tension at a given fiber length $L_F$, $F_{IO}$  s the maximum isometric force exerted at the optimum length of the muscle fibers, and $SK$ is a parameter specific for each muscle. Note that these parameters can be determined experimentally. 

An easy way to combine Eqs.(\ref{muscle:hill}) and (\ref{muscle:LT}) is to take the product of their normalized versions, as in \cite{Challis:1994,Battista:2015}. The resulting model is given by
\begin{equation}
\label{muscle:FVLT} F_{muscle}(L_F,V_F) = a_f \tilde{F}_{max} F_1(L_F) F_2(V_F),
\end{equation}
where $a_f$ is the activation strength of the muscle and $\tilde{F}_{max}$ is the normalized maximum isometric force generated at the full activation of the muscle fibers at their optimum lengths, and $F_1(L_F)$ and $F_2(V_F)$ are normalized versions of Eqs.(\ref{muscle:hill}) and (\ref{muscle:LT}), given by
\begin{align}
\label{muscle:F1} F_1(L_F) &= \exp\left[-\left( \frac{L_F/L_{FO}-1}{SK} \right)^2 \right], \\
\label{muscle:F2} F_2(V_F) &= \frac{1}{F_{max}} \left[ \frac{bF_{max}-a V_F}{b+V_F} \right].
\end{align}

%
%
%
%
\subsubsection{Muscle-Fluid-Structure Models: $3$-Element Hill Model}

\begin{figure}
    \centering
    \includegraphics[scale=0.27]{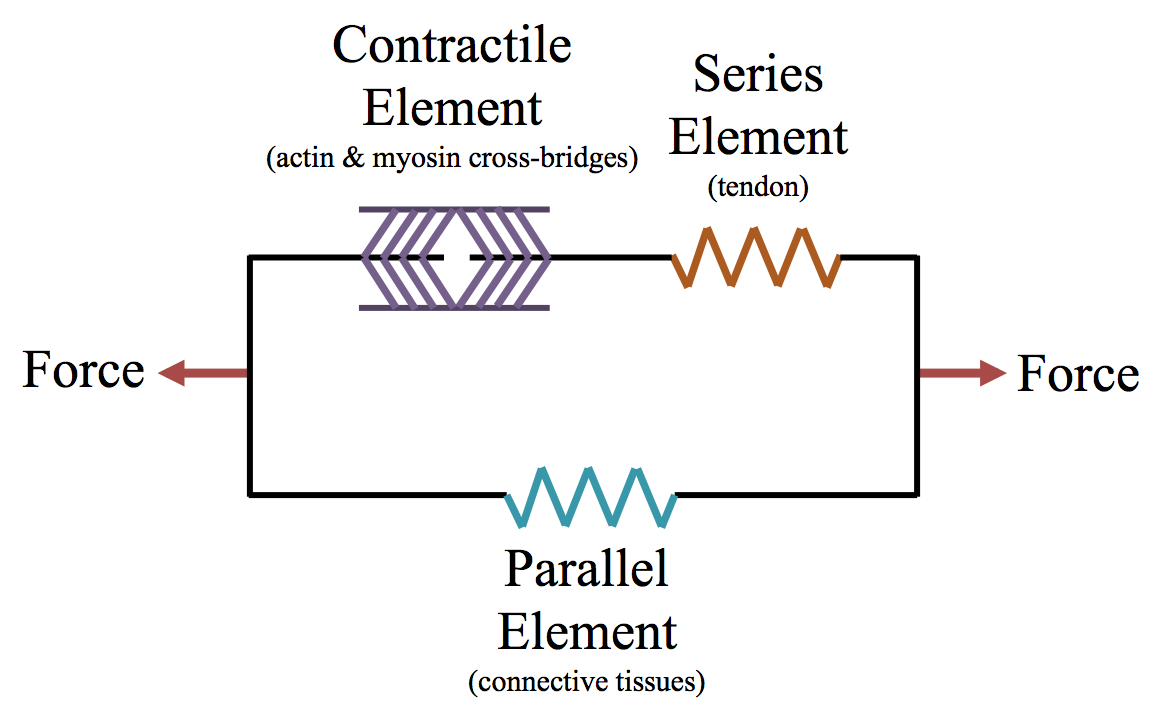}
    \caption{Schematic diagram of the $3$-Element Hill model of muscle contraction, containing a contractile element, series element, and parallel element modeling actin and myosin cross-bridges, tendons, and connective tissues, respectively.}
    \label{3elem:pic}
\end{figure}

The $3$-Element Hill model of muscle activation describes sustained muscle contraction by modeling the actin and myosin cross-bridges, muscle tendon, and connective tissues for a muscle. The model has a contractile element which models the force generated by the actin and myosin cross-bridges at the sarcomere level, and two non-linear spring elements, one in parallel and one in series with the contractile element.  The series element models the muscle tendon, i.e., the intrinsic elasticity of the myofilaments, and has a soft tissue response and provides energy storing mechanism. The parallel element takes care of the passive behavior when the muscle is stretched, representing connective tissues with a soft tissue-like behavior. Furthermore, the contractile element is fully extensible when inactive but capable of shortening when activated \cite{Hill:1938,Fung:1993}. The $3$-elements are depicted in Figure \ref{3elem:pic}.

The net force-length properties of a muscle are a result of both the active (contractile element and series element) and passive (parallel element) components' force-length characteristics. If $F_{CE}$, $F_{SE}$, and $F_{PE}$ represent the force produced by the contractile, series, and parallel elements respectively, their relations satisfy
\begin{align}
\label{3elem:force1} F_{tot} &= F_{SE}+F_{PE} \\
\label{3elem:force2} F_{CE} &= F_{SE},
\end{align}
where $F_{tot}$ is the total force produced by muscle contraction. Furthermore the relations for muscle shortening are
\begin{equation}
\label{3elem:length1} L_{tot} = L_{CE} + L_{SE} = L_{PE},
\end{equation}
where $L_{tot}$ is the total length of the muscle. Since the overall muscle length is conserved, if the series element is stretched, the contractile element must contract an equal amount.

As mentioned previously, to model the force produced from the series and parallel elements, we use non-linear springs, e.g.,
\begin{align}
\label{3elem:SE} F_{SE} &= k_{SE} \left( L-L_{CE} \right)^n \\
\label{3elem:PE} F_{PE} &= k_{PE} \left( L-L_{PE_{R}} \right)^n,
\end{align}
where $k_{SE}$ and $k_{PE}$ are the spring stiffnesses for the series and parallel elements respectively, and $L_{PE_{R}}$ is the resting-length of parallel element's non-linear spring. Note that the series element's spring has zero resting-length as it's length depends solely on the length of the contractile element. $n$ is an integer assumed to be greater than or equal to 2. There are many ways to represent these elements; these are only one possible choice. 

The contractile element assumes the length-tension and force-velocity relationship of muscle. For this reason, Eq.(\ref{muscle:FVLT}) is one possible choice for modeling its force generation, e.g.,
\begin{equation}
\label{3elem:CE} F_{CE} =  a_f \tilde{F}_{max} F_1(L_{CE}) F_2(V_{CE}),
\end{equation}
where $L_{CE}$ is the length of the muscle fibers and $V_{CE}$ is the contraction speed of the muscle fibers being represented in the contractile element. Another possible choice is described in \cite{Hamlet:2015}.

%
%

\subsection{Incorporated Models}
\label{IncorporatedModels}

In this section, we will describe inclusions within the current \textit{IB2d} software that are in addition to the fiber and material property models. Currently the following capabilities have been added to the framework:
\begin{enumerate}
\item Tracer particles
\item Concentration gradients of a chemical (advection-diffusion)
\item Background flow profiles (artificial forcing)
\item Basic electrophysiology frameworks
\end{enumerate}

These models are incorporated either by additional forcing terms in Eq. ($\ref{Navier_Stokes}$) (background flow profiles), a constitutive equation that depends on $u(\mathbf{x},t)$ without affecting the fluid momentum itself (tracers and concentration gradients), or by coupling into a specific fiber model (electrophysiology).

%
%
%
%
\subsubsection{Tracers}

Tracers are neutrally buoyant particles that move at the local fluid velocity. They have no impact on the fluid motion themselves, but rather ``go with the flow." If a tracer's position is given by $\mathbf{X}_{tr}$, their equation of motion is solely given by 
\begin{equation}
\label{tracer:eq} \frac{d\mathbf{X}_{tr}}{dt} = u(\mathbf{x},t), 
\end{equation}
where $u(\mathbf{x},t)$ is the background fluid velocity. Tracers are useful to observe the fluid motion during a simulation. The tracers move in the simulation by harnessing the discrete delta functions to interpolate the velocity of the exact position of the tracer.

%
%
%
%
\subsubsection{Concentration Gradients}

Similar to tracers, concentration gradients can be used to observe the motion of the underlying fluid; however, this model also incorporates diffusion. Rather than observing individual particles advect and diffuse, a concentration gradient is given as a continuum, $c$, which then gets spread out by an advection-diffusion constitutive equation,
\begin{equation}
\label{advdiff:eq} \frac{\partial c}{\partial t} + \mathbf{u}(\mathbf{x},t)\cdot \nabla c = D \Delta c,  
\end{equation}
where $D$ is the diffusivity coefficient. We note that $D$ is a constant in this formulation, there are no sources or sinks, and that $\mathbf{u}$ is, of course, assumed to be incompressible. The details of the numerical solver are found in \ref{appendix:advdiff}.

%
%
%
%
\subsubsection{Background Flow Profiles (Artificial Forcing)}

Although the computational domain is assumed to have periodic boundaries, one can induce a desired background flow profile by artificially adding a force onto the fluid, realized as an additional forcing term on Eq.(\ref{Navier_Stokes}). 

Essentially, the additional force will be a penalty-type term, which exerts a force onto the fluid grid, if the fluid velocity does not match the desired flow profile. Such a forcing term can take the form
\begin{equation}
\label{background:eq} \mathbf{F}_{arb} = k_{arb} \left( \mathbf{u}(\mathbf{x},t) - \mathbf{u}_{flow}(\mathbf{x},t) \right), 
\end{equation}
where $k_{arb}$ is the penalty-strength coefficient, and $\mathbf{u}_{flow}(\mathbf{x},t) $ is the desired background flow profile as in \cite{Santhanakrishnan:2009,Hamlet:2012}, where it was used to create parabolic inflow into a channel along the $x$-direction, i.e., 
\begin{equation}
\label{background:parabolic} \mathbf{u}_{flow}(\mathbf{x},t) = \left( \begin{array}{c}  U_{max}\left( 1 - \left( \frac{0.5-x}{d/2} \right)^2 \right) \\ 0  \end{array} \right). 
\end{equation}
We note this idea has also been used when a preferred mode of active force is desired \cite{Newren:2007,Teran:2009,Teran:2010}.

\begin{figure}
    \centering
    \includegraphics[width=0.5\textwidth]{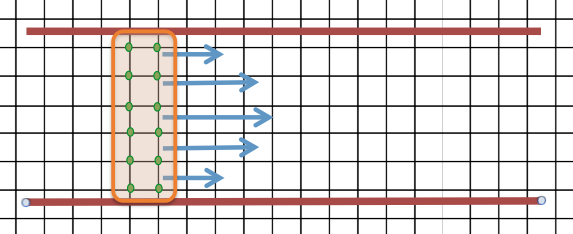}
    \caption{A depiction of exerting an arbitrary force onto the background fluid grid to obtain the desirable flow profile. The fluid grid is given by the rectangular grid with the selected grid points to enforce the penalty-force highlighted and circled in green and orange. The penalty force is applied to the fluid lattice points in green if the flow profiles do not match at those selected nodes. A cartoon rending of the resulting flow is illustrated as the blue arrows.}
    \label{IB_background_flow}
\end{figure}

This idea is illustrated in Figure \ref{IB_background_flow}, where the fluid grid is given by a rectangular grid with the selected grid points to enforce the penalty-force highlighted and circled in green and orange, respectively. The penalty force is applied to those green fluid lattice points if the flow profiles do not match at those selected nodes. A example rending of the resulting flow is illustrated as the blue arrows, if the desired flow profile is parabolic. Note that this addition of momentum (energy) into the system is not an issue, because of the assumed periodic boundary conditions.

%
%
%
%
\subsubsection{Electrophysiology}

A basic model of action potential propagation is incorporated using the FitzHugh-Nagumo equations (FHN) to model the system. FHN is a reduced order model of the Hodgkin-Huxley equations, which were the first equations to describe the propagation of an electrical signal along excitable cells. FHN has been incorporated into fluid-structure interaction models before \cite{Baird:2015}. The governing equations are given as
\begin{align}
\label{FHN:eq1} \frac{\partial v}{\partial t} &= \mathbb{D} \nabla v + v(v-a)(v-1) - w - \mathbb{I}(t) \\
\frac{\partial w}{\partial t} &= \epsilon(v-\gamma w),
\label{FHN:eq2}
\end{align}
where $v(x,t)$ is the membrane potential, $w(x,t)$ is the blocking mechanism, $\mathbb{D}$ is the diffusion rate of the membrane potential, $a$ is the threshold potential, $\gamma$ is the resetting rate, $\epsilon$ is the blocking strength parameter, and $\mathbb{I}(t)$ is an applied current, e.g., an initial stimulus potentially from pacemaker signal activation. Note that $v$ is the action potential and that $w$ can be thought to model a sodium blocking channel.

Once the membrane potential is found, it can be applied to the fiber model to activate pumping or initiate motion to which induces deformations of the structure to model desired biological or physical phenomena. Coupling the action potential to the generation of force can be done in many different ways for various applications, such as for cardiac contraction \cite{Griffith:2013,Baird:2015,Waldrop:2015} or locomotion \cite{Williams:1998,McMillen:2008,Tytell:2010}. 

%
%

\section{Work Flow}

We will now briefly describe the typical work flow for using the \textit{IB2d} software. Both MATLAB and Python have their own respective directories which in turn contain two folders: ``Examples" and ``IBM\textunderscore Blackbox". The Examples folder contains all currently implemented simulation examples including the necessary input files to run each simulation, with each example organized in its own folder. The IBM\textunderscore Blackbox folder contains all methods for solving the fluid-structure interaction problems. The software is set up such that the user will not have to change the underlying mechanics of the immersed boundary method unless they wish to make additions to the algorithm, e.g., implementing more fiber models, etc. 

\begin{figure}
    \centering
    \includegraphics[scale=0.52]{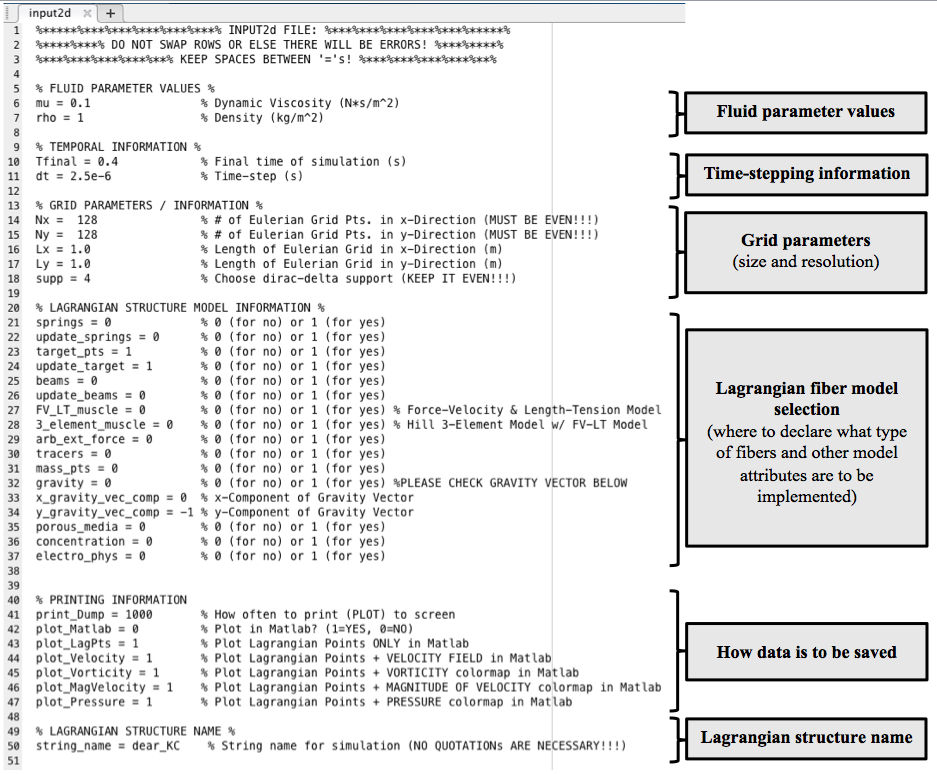}
    \caption{Descriptions of selections in \textit{input2d}. This file controls what inputs get passed to the main \textit{IB} driver method.}
    \label{workflow:input2d}
\end{figure}

Inside an example sub-directory, there are multiple files. Two files that must be in every example are \textit{input2d} and \textit{main2d}. \textit{input2d} is the file where the user chooses all parameters required for a simulation, i.e., the fluid parameters, temporal information, grid parameters, fiber model construction, how to save the data, etc. \textit{main2d} will read in this file and then read in the corresponding input files associated with the choices selected in \textit{input2d}. A graphical description of \textit{input2d} is given in Figure \ref{workflow:input2d}.

After setting desired parameters and selecting the necessary flags in \textit{input2d}, assuming the user has the appropriate associated input files corresponding to those selections the simulation is started by calling the \textit{main2d} script. This script reads in all the information from \textit{input2d} and passes it to the \textit{IBM\textunderscore Driver} script. Once the simulation finishes, a visualization folder, viz\textunderscore IB2d, will have all the Lagrangian structure and dynamical data from the simulation in .vtk format \cite{Schroder:2006}. .vtk files can be visualized using Paraview \cite{Paraview:2005} or VisIt \cite{HPV:VisIt}.

Each fiber model has an associated input file type, with the first line being the number of total fiber points associated with that type. If the immersed structure is called ``struct", the possible file types are as follows
\begin{itemize}
\item struct\textbf{.vertex}: A list of all the $(X,Y)$ initial coordinates of the Lagrangian points

%
%
\begin{figure}[H]
    \centering
    \includegraphics[scale=0.45]{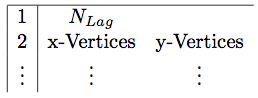}
    \caption{Input format for the .vertex file}
    \label{workflow:vertexfile}
\end{figure}

%
%
\item struct\textbf{.spring}: A list of the master and slave nodes for each linear spring along with their associated spring stiffness, resting-length, and degree of non-linearity. Note that if using only Hookean springs, the degree of non-linearity can be omitted and \textit{IB2d} will automatically assume linear springs.

\begin{figure}[H]
    \centering
    \includegraphics[scale=0.59]{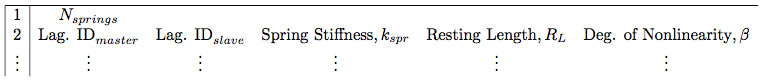}
    \caption{Input format for a .spring file}
    \label{workflow:springfile}
\end{figure}

%
%
\item struct\textbf{.beam}: A list of the left, middle, and right Lagrangian indices associated with each torsional spring (beam) and their associated beam stiffness and curvature. 

\begin{figure}[H]
    \centering
    \includegraphics[scale=0.45]{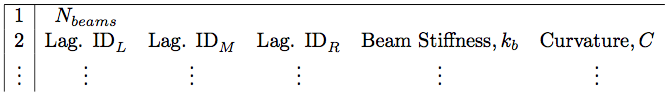}
    \caption{Input format for a .beam (torsional spring) file}
    \label{workflow:beamfile}
\end{figure}

%
%
\item struct\textbf{.target}:A list of all target point indices with their associated target point stiffness.

\begin{figure}[H]
    \centering
    \includegraphics[scale=0.45]{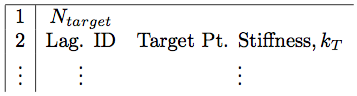}
    \caption{Input format for a .target file}
    \label{workflow:targetfile}
\end{figure}

%
%
\item struct\textbf{.mass}: A list of all Lagrangian mass point indices along with their associated mass-spring stiffness and mass. 

\begin{figure}[H]
    \centering
    \includegraphics[scale=0.45]{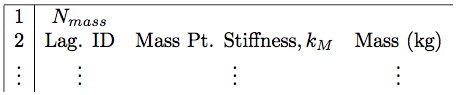}
    \caption{Input format for a .mass file}
    \label{workflow:massfile}
\end{figure}

%
%
\item struct\textbf{.porous}: A list of all porous Lagrangian points, along with their associated porosity coefficient, $\alpha$, and their Stencil ID.

\begin{figure}[H]
    \centering
    \includegraphics[scale=0.39]{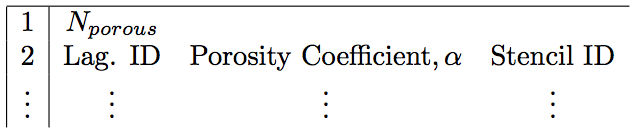}
    \caption{Input format for a .porous file}
    \label{workflow:porousfile}
\end{figure}

\begin{figure}[H]
    \begin{subfigure}{0.4\textwidth}
        \centering
        \includegraphics[width=0.8\textwidth]{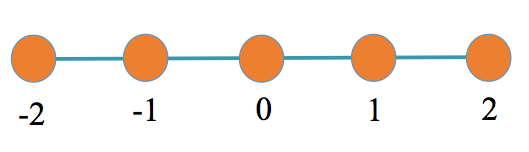}
        \caption{}
        \label{fig:por1} 
    \end{subfigure} 
    \begin{subfigure}{0.6\textwidth}
        \centering
        \includegraphics[width=0.75\textwidth]{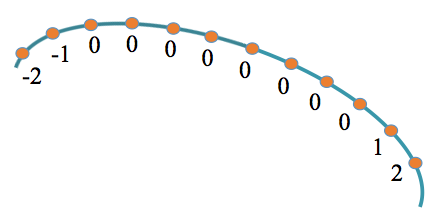}
        \caption{}
        \label{fig:por2}
    \end{subfigure}
\caption{(a) Order of porous stencil IDs. (b) An example of how the stencil IDs are defined using the porous structure from Figure \ref{IB_Porous}. }
\label{Porous_Stencil}
\end{figure}

Note: the stencil ID is an integer between $\{-2,-1,0,1,2\}$, which declares which points around the node of interest will be used in the derivative calculations. For porosity, you need a minimum of $4$ nodes surrounding the porous node. This idea is illustrated in Figure \ref{Porous_Stencil}, which shows how a group of $5$ porous stencil IDs would be labeled and an example of how they would be labeled in practice for a toy example.

%
%
\item struct\textbf{.muscle}: A list of all $FV$-$LT$ muscle master and slave node Lagrangian indices, along with their associated muscle length in which the fibers generate their maximum tension ($L_{FO}$), $SK$-parameter, thermodynamic parameters, $a$ and $b$, respectively, and the muscle's maximum load at zero contractile velocity ($F_{max}$). 

\begin{figure}[H]
    \centering
    \includegraphics[scale=0.45]{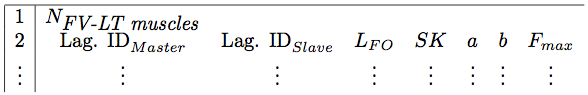}
    \caption{Input format for a .muscle (FV-LT muscle) file}
    \label{workflow:FLTVfile}
\end{figure}

%
%
\item struct\textbf{.tracer}: A list of all tracer particles' initial coordinates, $(X,Y)$. 
\begin{figure}[H]
    \centering
    \includegraphics[scale=0.45]{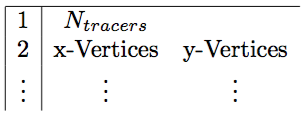}
    \caption{Input format for a .tracer  file}
    \label{workflow:tracers}
\end{figure}

\end{itemize}

These file formats are consistent with those necessary to run simulations in IBAMR, making this software an appropriate learning and analysis tool before scaling up to larger and more highly resolved simulations.

%
%

\section{Examples}

In this section we will present some examples which show some of the software's functionality. The software currently contains over $30$ examples built-in; we will choose some that highlight specific features of the software. 

\begin{itemize}
    \item The Rubber-band
    \item The Flexible Beam
    \item The ``Date" 
    \item Falling Spheres under Gravity vs. Pulsatile Flow
\end{itemize}


\subsection{The Rubber-band}

\begin{figure}[H]
    \centering
    \includegraphics[scale=0.4]{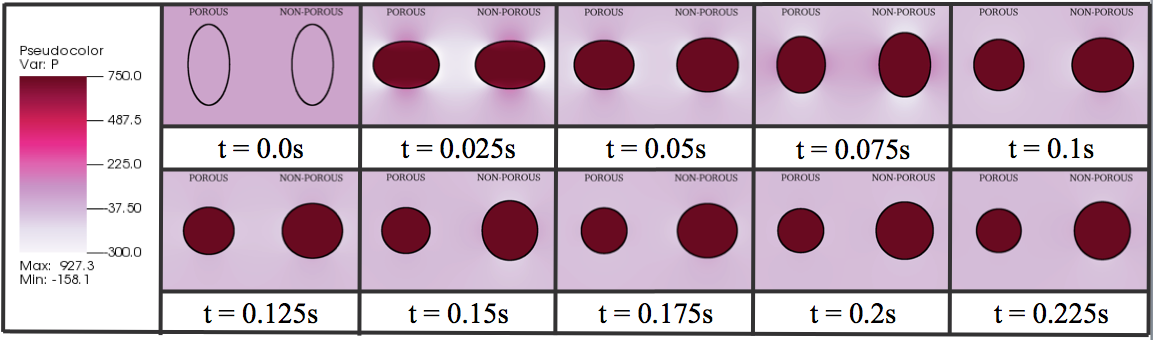}
    \caption{A comparison of a porous rubberband (left) and a non-porous rubberband (right). The colormap is of the fluid pressure. It is clear that both rubberbands start stretched from their equilibrium position, but end at a circle; however, the porous rubberband does not conserve the same volume, as fluid flows through it as the simulation progresses.}
    \label{Example:Porous_Rubberband}
\end{figure}

The rubber-band is one of the quintessential problems in FSI. The band is composed of springs between adjacent nodes, all with a preferred resting length of zero and constant spring stiffness. We will model two such rubber-bands with equivalently perturbed initial states, but the rubber-band on the left will be porous at each Lagrangian node to show how fluctuations cause it to lose volume over time. In summary, the fiber models used are:

\begin{itemize}
    \item Linear Springs
    \item Porosity (non-traditional rubber-band)
\end{itemize}

The simulation starts with the rubber band stretched into an elliptical shape with a fixed volume of fluid trapped within the elastic band. Since the resting length is zero, the rubber bands will be driven toward the lowest state of energy that minimizes length for a given internal volume, i.e., a circle. As the band moves toward this equilibrium position, it will contract and expand periodically across the semi- major and semi-minor directions of the axis. In the case of the porous rubber-band, it will also shrink in size. Simulation images are shown in Figure \ref{Example:Porous_Rubberband}.



\subsection{The Flexible Beam}

\begin{figure}[H]
    \centering
    \includegraphics[scale=0.41]{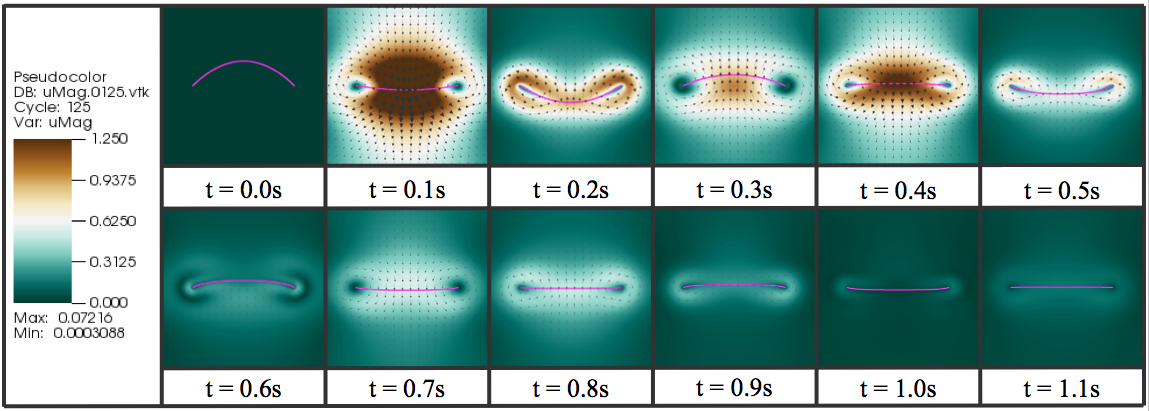}
    \caption{The flexible beam shown at various times during the simulation. The colormap is of the magnitude of velocity is depicted along with the background fluid velocity vectors.}
    \label{Example:FlexibleBeam}
\end{figure}

The flexible beam is another standard problem in FSI. It includes a ``beam" composed of adjacent torsional springs between three successive nodes and tethered to two fixed points at the ends of the beam, modeled using target points. The torsional springs all have a preferred `curvature' of zero, making any perturbation in the geometry move towards a straight line. The fiber models implemented are: 

\begin{itemize}
    \item Torsional Springs
    \item Target Points (fixed)
\end{itemize}

The simulation starts with the beam having been perturbed from its equilibrium position by an ellipsoidal arc. Since the preferred torsional spring `curvature' is zero, the torsional springs will move the system towards an equilibrium where they all line up. Since the ends of the beam are fixed horizontally from each other, the lowest energy state of the system will be when the beam forms a horizontal line between the two end points. Time-slices from the simulation are illustrated in Figure \ref{Example:FlexibleBeam}.


\subsection{The ``Date"}
\label{The_Date_Example}

\begin{figure}[H]
    \centering
    \includegraphics[scale=0.41]{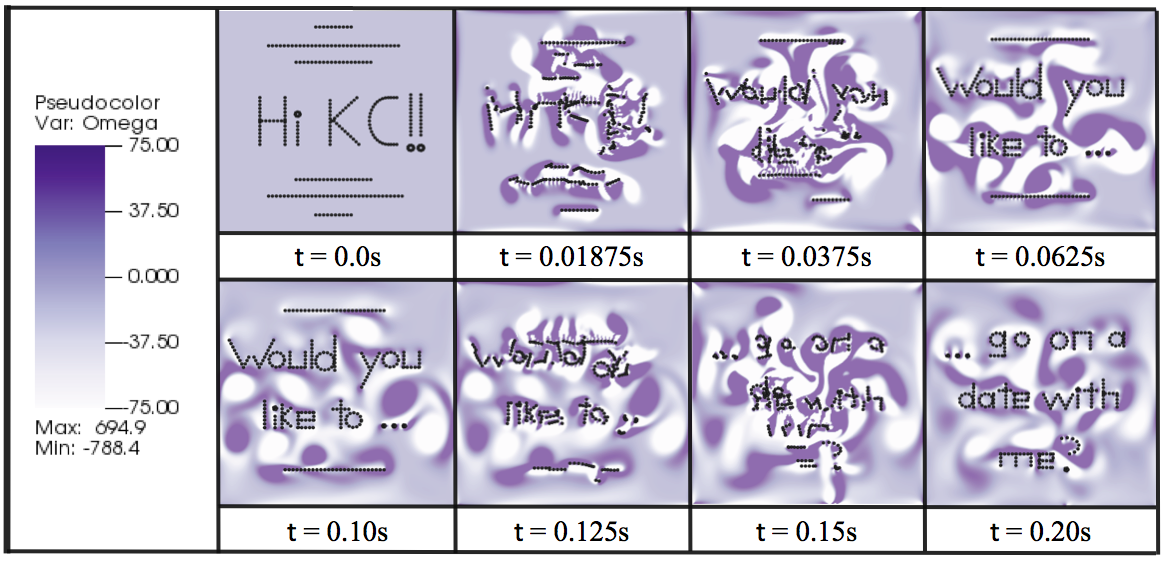}
    \caption{``The Date" example shows target points moving around the computational domain by interpolating positions between three states which spell out certain phrases. The background color map is vorticity.}
    \label{Example:TheDate}
\end{figure}

The ``\emph{Date}" illustrates the software's ability to update target point positions. Every Lagrangian point in this simulation is modeled as a target point and is given a specific location. As the simulation progresses, Lagrangian point positions are moved to new prescribed positions via an interpolation function, implemented within \emph{update$\_$target$\_$point$\_$positions}. The fiber models used are:

\begin{itemize}
    \item Target Points (with dynamically updating positions)
\end{itemize}

The simulation starts in a configuration that spells out ``Hi KC!!" enclosed within a square of Lagrangian points with a few other lines of Lagrangian points in decor. As the simulation progresses, those Lagrangian points are interpolated to their next prescribed configuration, spelling ``Would you like to\ldots". Finally, they move to their final configuration, spelling ``\ldots go on a date with me?" Snapshots of the simulation are illustrated in Figure \ref{Example:TheDate}.

As stated before, the script \emph{update$\_$target$\_$point$\_$positions} was used to dynamically update the position of the target points. Furthermore, within this script, one could also change any of the target point parameters, i.e., target point tethering stiffnesses or position. 

Similarly, other scripts can be used to dynamically update parameters of springs and/or torsional springs as a simulation progresses, e.g., `\emph{update$\_$springs} and `\emph{update$\_$beams}'. In the case of springs, one can update the spring stiffnesses, resting-lengths, or even non-linearity properties (\emph{see Example$\_$HeartTube}), and in the case of torsional springs one can update their torsional spring stiffness or preferred `curvature' (\emph{See Example$\_$Jellyfish}).


\subsection{Falling Spheres under Gravity vs. Pulsatile Flow}

\begin{figure}[H]
    \centering
    \includegraphics[scale=0.37]{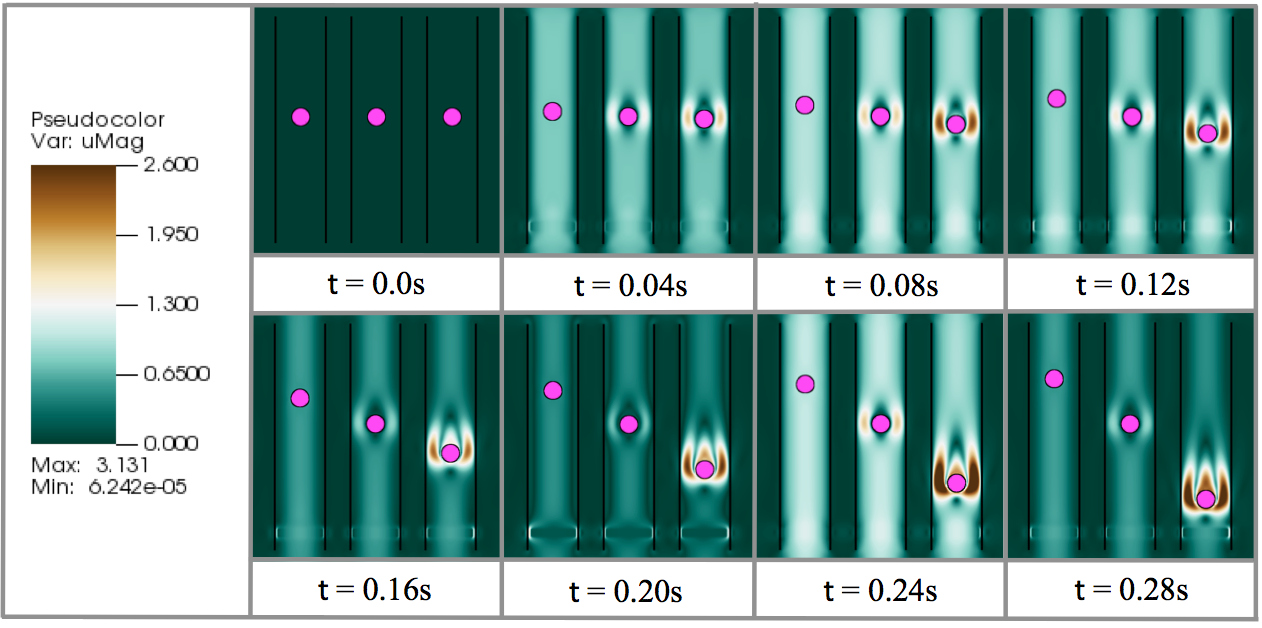}
    \caption{3 spheres of different masses, with each sphere composed of uniform mass points, under the influence of gravity with a pulsatile flow competing against gravity upwards. The sphere on the left is the lightest and sphere on the right is the heaviest. In the left case, the pulsatile flow dominates, while on the right gravity dominates, and in the middle, the pulsatile flow and gravitational forces are approximately equivalent. The colormap depicts magnitude of velocity.}
    \label{Example:MassGravityRace}
\end{figure}

This example simulates a competition between spheres falling under gravity and upward flow, which acts to help the spheres resist gravity. There are three vertical channels composed of fixed target points. In each channel there is also a sphere, composed of stiff springs and stiff torsional beams between each adjacent Lagrangian node. Each point on the sphere has an associated mass with gravity pointed in the downward vertical direction. There is also a net prescribed flow upwards, arising from an artificial force directly applied onto the Eulerian (fluid) grid. The fiber models and functionality used are:

\begin{itemize}
    \item Linear Springs
    \item Torsional Springs
    \item Target Points (fixed)
    \item Massive Points
    \item Artificial Forcing on the Fluid Grid
\end{itemize}

The simulation begins with three spheres of different masses. Each sphere is itself composed of individual uniform mass points; however, the individual mass points differ from sphere to sphere. Gravity is acting on the masses to pull them downward while upward flow is providing a force in the opposite direction. In one case, the flow dominates, in another gravity is balanced by the imposed flow, and in the other case, gravity is dominate. This is illustrated in Figure \ref{Example:MassGravityRace}.


\subsection{Idealized Swimmer}

\begin{figure}[H]
    \centering
    \includegraphics[width=0.98\textwidth]{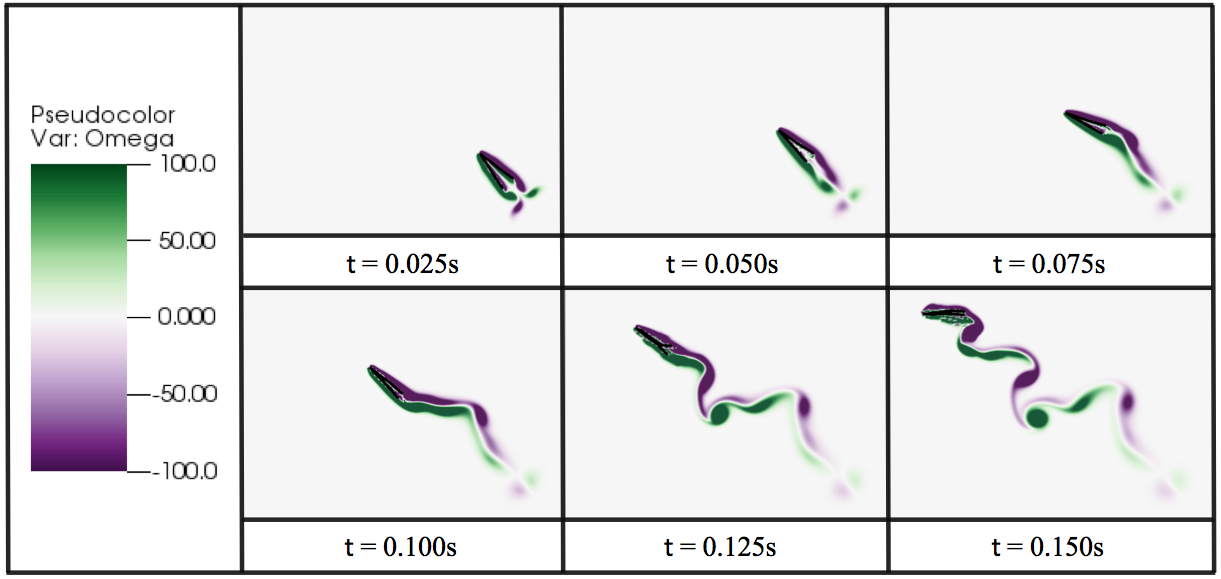}
    \caption{An idealized swimmer moving forward and turning due to the asynchronous muscle activation. The colormap illustrates vorticity.}
    \label{Example:Swimmer}
\end{figure}

This example uses the $3$-element Hill muscle model to cause an idealized swimmer, shaped like a $V$, to move forward and turn. There are stiff linear springs and stiff torsional springs connecting all adjacent Lagrangian points. Only $3$ muscles connect one leg of the $V$ to the other, and are equally spaced at intervals $3L/10, 2L/5$, and $9L/10$ down the leg of the swimmer, where $L$ is the length of each leg.   

\begin{itemize}
    \item Linear and Non-Linear Springs
    \item Torsional Springs
    \item 3-Element Hill Model
\end{itemize}

The simulation begins with the swimmer in a $V$-shaped starting position at rest. Throughout the simulation the muscles fire out of phase, causing the swimmer to move forward and turn. The swimming behavior is shown in Figure \ref{Example:Swimmer}.

%
%

\section{Code Validation}

In this section we present a validation of the code, both in the form of a convergence study for a particular example, as well as in the form of a comparison to experimental data. 

\begin{figure}[H]
    \centering
    \includegraphics[width=\textwidth]{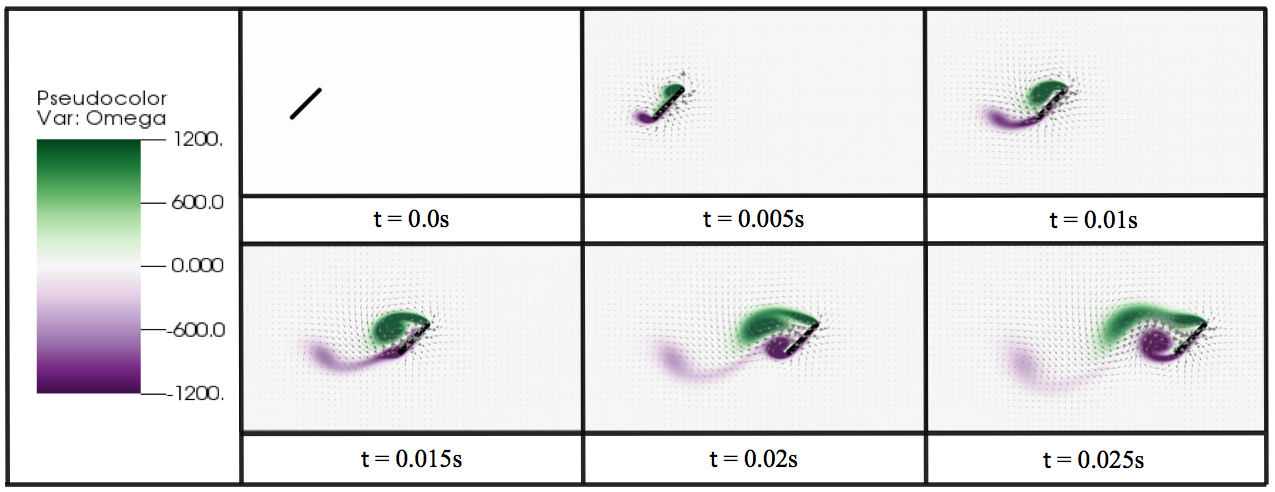}
    \caption{A cross-section of an insect wing moving laterally from left to right in a prescribed manner for $Re=128$.The background colormap depicts vorticity and the vector field is the fluid velocity.}
    \label{Example:Wing}
\end{figure}

Consider the example of a cross-sectional piece of an insect wing moving laterally across the domain at a $45$ degree angle of attack at $Re=128$. The insect wing's motion is governed by updating the target point positions, as in the example in Section \ref{The_Date_Example}. Snapshots from the simulation are shown in Figure \ref{Example:Wing}. 

%
%

\subsection{Convergence Study}

In this section, we perform a convergence study focusing on the forces in the $x$- and $y$-directions, respectively referred to as drag and lift, which act on the immersed structure (wing). We ran the simulations for different grid resolutions of the fluid domain (and complementary Lagrangian spacing in the immersed structure) with $Re=128$ at equivalent time-step, $dt$, and uniform material properties of the wing. The fluid grid resolutions studied were $\{32\mbox{x}32, 64\mbox{x}64, 96\mbox{x}96, 128\mbox{x}128, 256\mbox{x}256, 512\mbox{x}512, 768\mbox{x}768, 1024\mbox{x}1024\}$. The forces over time are plotted below in Figure \ref{IB:Insect_Force_Time}.

\begin{figure}[H]
    \begin{subfigure}{0.5\textwidth}
        \centering
        \includegraphics[width=0.98\textwidth]{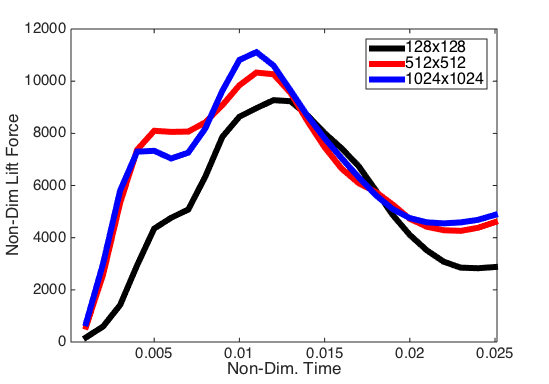}
        \caption{}
        \label{IB_Insect_Lift}
    \end{subfigure}
    \begin{subfigure}{0.5\textwidth}
        \centering
        \includegraphics[width=0.98\textwidth]{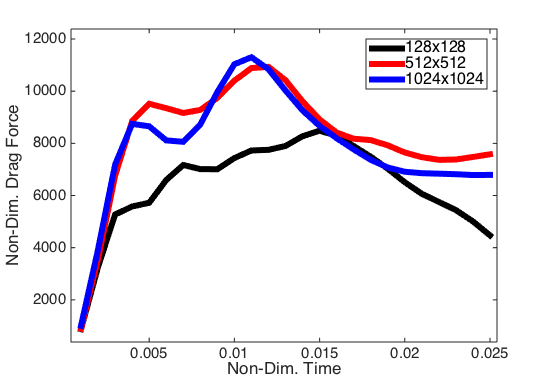}
        \caption{}
        \label{IB_Insect_Drag}
    \end{subfigure}\\ \\
\caption{Non-dimensional lift (\ref{IB_Insect_Lift}) and drag (\ref{IB_Insect_Drag}) forces vs non-dimensional time for a cross-section of an insect wing moving laterally at $Re=128$.}
\label{IB:Insect_Force_Time}
\end{figure}

The mean lift and drag forces were calculated over the wing at $t=0.025s$ for each simulation, and then the relative error was computed between each simulation and a highly resolved case using IBAMR with $1024\mbox{x}1024$ fluid grid resolution.

\begin{figure}[H]
    \centering
    \includegraphics[width=0.65\textwidth]{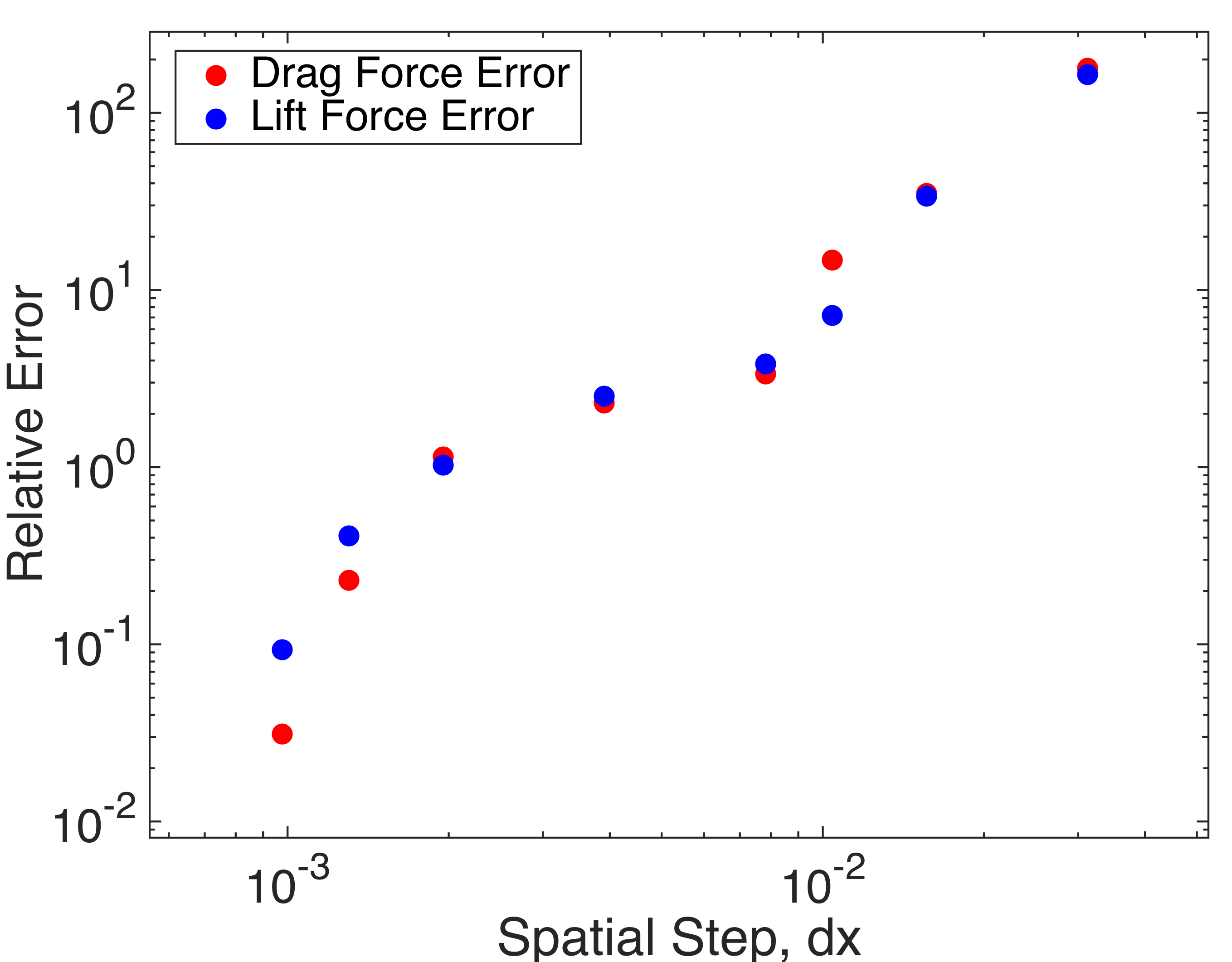}
    \caption{A convergence study of the relative error of the lift and drag force, between each simulation and the highly resolved simulation using IBAMR with $1024\mbox{x}1024$ resolution on the fluid grid. We note that the horizontal axis is the spatial step size, dx, where $dx = 1/N$, and $N=\{32,64,96,128,256,512,768,1024\}$.}
    \label{IB:Conv_Study}
\end{figure}

Applying a best fit line to the data produces the relative error convergence rates
\begin{align}
\label{drag_err} \mbox{Relative Error}_{\mbox{ Drag}} &\sim 281838.29\ (dx)^{2.16} \\
\label{lift_err} \mbox{Relative Error}_{\mbox{ Lift}} &\sim 67920.36\ (dx)^{1.83}.
\end{align}

%
%

\subsection{Experimental Validation}

In this section we compare simulation results from \textit{IB2d} to experimental data for a cross-section of an insect wing moving laterally across the domain for three orders of magnitude of $Re$. The experimental data was obtained using particle image velocimetry (PIV) \cite{Kiger:2016}, using a dynamically-scaled flapping robot, e.g., Robofly. \cite{Dickinson:1993,Birch:2004}. The simulations were run on a $1024$x$1024$ grid.

\begin{figure}[H]
    \centering
    \includegraphics[width=0.8\textwidth]{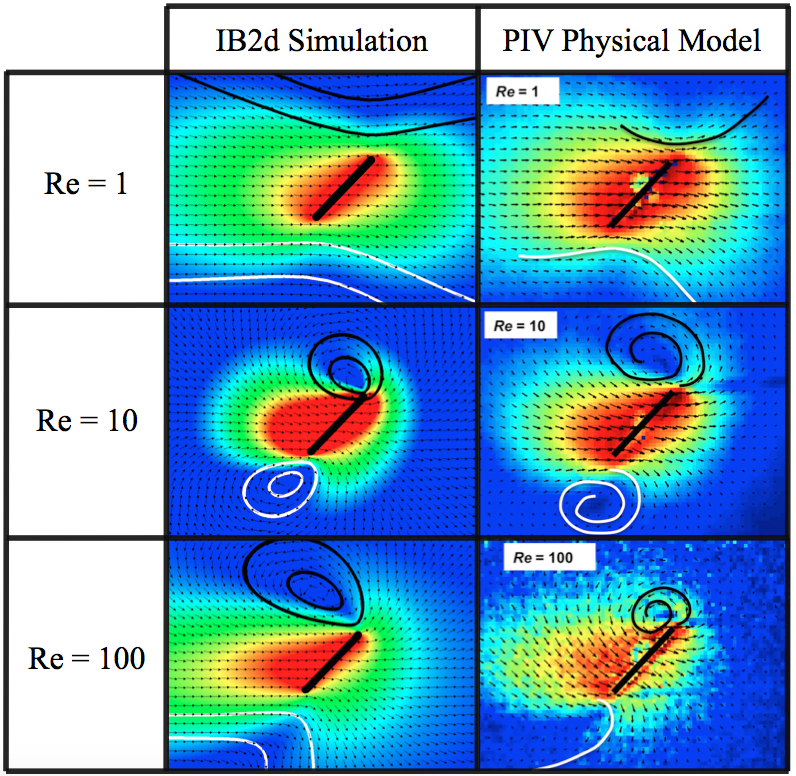}
    \caption{Comparison of \textit{IB2d} simulation snapshots and PIV experimental data for a wing moving laterally across the domain for $Re=\{1,10,100\}$. The figures show the magnitude of velocity, background velocity field, and streamlines.}
    \label{IB:Exp_Validation}
\end{figure}

Figure \ref{IB:Exp_Validation} shows a comparison of snapshots taken from \textit{IB2d} and the PIV physical model over a range of $Re$. The basic flow structures are reproduced in all cases. 


%
%

\section{Discussion and Conclusion}

\textit{IB2d} is immersed boundary software with full implementations in both MATLAB and Python 3.5. It offers a vast array of fiber model options for constructing the immersed structure and has functionality for advection-diffusion, artificial forcing, muscle mechanics, and electrophysiology. Furthermore, having been written in high-level programming languages, it allows one to implement new fiber models and functionality easily and at an accelerated rate. 

High-level programming languages also come with a few drawbacks. Grid sizes should not be implemented beyond a $512\times 512$ resolution due to computational costs. If higher resolution is required, we suggest moving to IBAMR. Additionally unlike IBAMR, \textit{IB2d} was strictly designed for $2D$ applications. While full $3D$ simulations are often desired, some applications may only require fluids with two-dimensions \cite{Tytell:2010,Crowl:2011,Zhang:2014,Lewis:2015,Waldrop:2015}. \textit{IB2d} was written in $2D$ to make it more readable and to lend itself for easier modification, particularly as a first step in trying to implement a new model. 

The format of \textit{IB2d} was designed to mirror the input file formats used in IBAMR, and as such can used as a stepping stone to using IBAMR. Neither \textit{IB2d} nor IBAMR include functionality for compressible fluids, non-Newtonian fluids, or variable density fluid applications at this time, but there are plans to incorporate them in the future. 

At this time neither \textit{IB2d} nor IBAMR include a turbulence model for large $Re$ simulations. For these applications, one is directed to use other software packages such as OpenFOAM by OpenCFD LTD \cite{openFOAM:openFOAM}, which is capable of FSI applications and is open source. Commercial software, such as COMSOL \cite{COMSOL:2011} and ANSYS Fluent \cite{Fluent:2016} can model FSI as well. OpenFOAM, COMSOL, and Fluent allow easier entry into FSI through well developed GUIs and manuals. However, licenses for COMSOL and Fluent are both expensive when not being used for academic teaching purposes. It is also difficult to implement or modify numerical approaches in COMSOL or Fluent, and there would be a steep learning curve for OpenFOAM. 

Note that there are other methods for simulating fluid-structure interactions in addition to Peskin's  immersed boundary method. Some examples include immersed interface methods \cite{Lee:2003,Li:2003}, sharp interface methods \cite{Ubbink:1999,Udaykumar:2001}, the blob projection method \cite{Cortez:2000}, and level set methods \cite{Sethian:1999,Osher:2002}. These methods have the benefit that they can capture high resolution of flow near interface when desired. However, the authors are not aware of any open source implementations at this time, and thus they require a large entry time for research and development - especially in the case for $3D$, adaptive, or parallelized applications. The mathematical work for compressible, non-Newtonian, and variable density fluids applications may be limited at this time as well. Furthermore, most sharp interface approaches have been limited to thin structures (e.g., elastic membranes) or rigid bodies. 

For teaching FSI applications, or fast implementations of new fiber models, numerical models and approaches, or varying fluid solvers, \textit{IB2d} is an ideal environment.


%
%

\section{Acknowledgements}
The authors would like to thank Charles Peskin for the development of immersed boundary method and Boyce Griffith for IBAMR, to which many of the input files structures of \textit{IB2d} are based. We would also like to thank Austin Baird, Christina Battista, Robert Booth, Namdi Brandon, Kathleen Carroll, Christina Hamlet, Alexander Hoover, Shannon Jones, Andrea Lane, Julia Samson, Arvind Santhanakrishnan, Michael Senter, Anne Talkington, and Lindsay Waldrop for comments on the design of the software and suggestions for examples. This project was funded by NSF DMS CAREER \#1151478, NSF CBET \#1511427
NSF DMS \#1151478, NSF POLS \#1505061 awarded to L.A.M. Funding for N.A.B. was provided from an National Institutes of Health T32 grant [HL069768-14; PI, Christopher Mack] and UNC Dissertation Completion Fellowship.

%
%

\appendix

%
%
%
%
\section{Discretization}
\label{appendix-discretization}

The discretizations used in \textit{IB2d} for solving The Navier-Stokes equations, e.g., (\ref{Navier_Stokes}) and (\ref{Incompressibility}), for computing normal derivatives for porous elements, and for advection-diffusion will be described below. 

%
%
%
%
\subsection{Discretizing the Navier-Stokes Equations}
\label{appendix:fluidsolve}
$ $\\
\indent \textit{IB2d} uses finite difference approximations to discretize the Navier-Stokes equations on a fixed lattice, e.g., the Eulerian (fluid) grid. It follows the discretization described in \cite{Peskin:1996,Peskin:2002}, and are implicitly defined as follows

\begin{align}
\label{Discrete:NS} \rho \Bigg( \frac{ \textbf{u}^{k+1} - \textbf{u}^{k} }{\Delta t} + S_{\Delta x} \left( \textbf{u}^k \right) \textbf{u}^{k} \Bigg) - \textbf{D}^0 &p^{k+1} = \mu \sum_{\alpha=1}^2 D_{\alpha}^{+} D_{\alpha}^{-} \textbf{u}^{k+1} + \textbf{F}^k \\ \nonumber \\
\label{Discrete:Incompressibility} \textbf{D}^0 \cdot \textbf{u}^{k+1} &= 0,
\end{align}

where $\Delta t$ and $\Delta x$ are the time-step and Eulerian meshwidth, respectively, and $\rho$ and $\mu$ are the density of the and kinematic viscosity of the fluid, respectively. $\textbf{D}^0$ is the central differencing operator, defined as

\begin{equation}
\label{Discrete:CentralDiff_1} \textbf{D}^0 = \left( D_1^0, D_2^0 \right),
\end{equation}

with 

\begin{equation}
\label{Discrete:CentralDiff_2} \left(D_{\alpha}^0\phi\right)(\textbf{x}) = \frac{ \phi\big( \textbf{x} + \Delta \textbf{x} e_{\alpha} \big) - \phi\big( \textbf{x} - \Delta\textbf{x} e_{\alpha} \big) }{2\Delta x},
\end{equation}

where $\left(e_1,e_2\right)$ is the standard basis in $\mathbb{R}^2.$ The viscous term, given by $ \sum_{\alpha=1}^2 D_{\alpha}^{+} D_{\alpha}^{-} \textbf{u}^{k+1}$, is a difference approximation to the Laplacian, where the $D_{\alpha}^{\pm}$ operators are the forward and backward approximations to $\frac{ \partial}{\partial x_{\alpha} }.$ They are defined as

\begin{align}
\label{Discrete:Forward} \left (D_{\alpha}^+\phi\right)(\textbf{x}) &= \frac{ \phi\big( \textbf{x} + \Delta \textbf{x} e_{\alpha} \big) - \phi\big( \textbf{x} \big) }{\Delta x} \\ \nonumber \\
\label{Discrete:Backward} \left (D_{\alpha}^{-}\phi\right)(\textbf{x}) &= \frac{ \phi\big( \textbf{x} \big) - \phi\big( \textbf{x} - \Delta\textbf{x} e_{\alpha} \big) }{\Delta x}. 
\end{align}

The skew-symmetric difference operator, $S_{\Delta x}$, serves as an approximation to the non-linear advection term, $\textbf{u}\cdot \nabla\textbf{u}$, and is defined as follows

\begin{equation}
\label{Discrete:Skew} S_{\Delta x} = \frac{1}{2} \left[ \textbf{u}\cdot \textbf{D}_{\Delta x}^0\phi  + \textbf{D}_{\Delta x}^0\phi\cdot (\textbf{u}\phi)  \right]. 
\end{equation}

Using the discretizations (\ref{Discrete:CentralDiff_2}), (\ref{Discrete:Forward}), (\ref{Discrete:Backward}), and (\ref{Discrete:Skew}), the equations (\ref{Discrete:NS}) and (\ref{Discrete:Incompressibility}) are linear in $\textbf{u}^{k+1}$ and $p^{k+1}$. To solve for $\textbf{u}^{k+1}$ and $p^{k+1}$ from $\textbf{u}^{k}, p^{k+1},$ and $F^{k}$, the Fast Fourier Transform (FFT) was implemented \cite{Cooley:1965,Press:1992}. Note that this assumes a periodic domain. Future implementations will include non-square domains and projection methods to incorporate Dirichlet or Neumann Boundary conditions \cite{Chorin:1967,Brown:2001}. 

The Navier-Stokes equations need not be discretized in this manner, and this is where one could implement a fluid solver and discretization of their choice, e.g., finite element or Lattice Boltzmann \cite{Zhu:2011}. However, further consideration must be taken into account on how to spread the Lagrangian forces to the Eulerian grid and move the Lagrangian structure at the local fluid velocity, i.e., Eqns.(\ref{IBM_Force}) and (\ref{IBM_Velocity}), respectively, if implementing in \textit{IB2d}.

%
%
%
%
\subsection{Discretizing the Normal Derivatives on the Boundary}
\label{appendix:porous}
$ $\\
\indent The normal vector to the Lagrangian structure is given by \cite{Kim:2006,Stockie:2009},
\begin{equation}
\label{Discrete:Normal} \textbf{n} = \tau\times e_3,
\end{equation}
where 
\begin{equation}
\label{Discrete:Tangent} \tau = \frac{ \frac{\partial \textbf{X} }{\partial s } }{ \left| \frac{\partial \textbf{X} }{\partial s } \right| }
\end{equation}

Hence we have that 
\begin{equation}
\label{Discrete:NormalVec} \textbf{n} = \left( \frac{ \partial Y / \partial s }{ \left| \partial \textbf{X} / \partial s  \right| } , - \frac{ \partial X / \partial s }{ \left| \partial \textbf{X} / \partial s  \right| }   \right).
\end{equation}

Unlike \cite{Kim:2006}, who used a $3$-$pt$ central differencing operator to compute $\frac{\partial \textbf{X} }{\partial x }$, we compute the partial derivatives using a $5$-$pt$ differentiation stencil. We do this to both minimize error near end-points of a porous structure and allow functionality for non-closed porous structures. Hence we implement the following five different differentiation operators,

\begin{align*}
(D_{\alpha}^{5,-2}\boldsymbol{\phi})(s) &= \frac{ -\frac{25}{12}\boldsymbol{\phi}\big(s-2\Delta s e_{\alpha}\big)  + 4\boldsymbol{\phi}\big(s-\Delta s e_{\alpha}\big)  -3\boldsymbol{\phi}\big(s\big) + \frac{4}{3}\boldsymbol{\phi}\big(s+\Delta s e_{\alpha}\big)  -\frac{1}{4} \boldsymbol{\phi}\big(s+2\Delta s e_{\alpha}\big)                }{\Delta s} \\
(D_{\alpha}^{5,-1}\boldsymbol{\phi})(s) &= \frac{ -\frac{1}{4}\boldsymbol{\phi}\big(s-2\Delta s e_{\alpha}\big)   -\frac{5}{6}\boldsymbol{\phi}\big(s-\Delta s e_{\alpha}\big)  + \frac{3}{2}\boldsymbol{\phi}\big(s\big) -\frac{1}{2} \boldsymbol{\phi}\big(s+\Delta s e_{\alpha}\big)  + \frac{1}{12}\boldsymbol{\phi}\big(s+2\Delta s e_{\alpha}\big)                }{\Delta s} \\
(D_{\alpha}^{5,0}\boldsymbol{\phi})(s) &= \frac{ \frac{1}{12}\boldsymbol{\phi}\big(s-2\Delta s e_{\alpha}\big)  - \frac{2}{3}\boldsymbol{\phi}\big(s-\Delta s e_{\alpha}\big)  + \frac{2}{3} \boldsymbol{\phi}\big(s+\Delta s e_{\alpha}\big)  - \frac{1}{12}\boldsymbol{\phi}\big(s+2\Delta s e_{\alpha}\big)                }{\Delta s} \\
(D_{\alpha}^{5,1}\boldsymbol{\phi})(s) &= \frac{ -\frac{1}{12}\boldsymbol{\phi}\big(s-2\Delta s e_{\alpha}\big)  + \frac{1}{2}\boldsymbol{\phi}\big(s-\Delta s e_{\alpha}\big)  - \frac{3}{2}\boldsymbol{\phi}\big(s\big) + \frac{5}{6}\boldsymbol{\phi}\big(s+\Delta s e_{\alpha}\big)  + \frac{1}{4}\boldsymbol{\phi}\big(s+2\Delta s e_{\alpha}\big)                }{\Delta s} \\
(D_{\alpha}^{5,2}\boldsymbol{\phi})(s) &= \frac{ \frac{1}{4}\boldsymbol{\phi}\big(s-2\Delta s e_{\alpha}\big)  - \frac{4}{3}\boldsymbol{\phi}\big(s-\Delta s e_{\alpha}\big)  + 3\boldsymbol{\phi}\big(s\big) - 4\boldsymbol{\phi}\big(s+\Delta s e_{\alpha}\big)  + \frac{25}{12}\boldsymbol{\phi}\big(s+2\Delta s e_{\alpha}\big)                }{\Delta s} \\
\end{align*}

%
%
%
%
\subsection{Discretizing the Advection-Diffusion Equation}
\label{appendix:advdiff}
$ $\\
The concentration is discretized on the same resolution as the Eulerian grid. Each nodal point has a scalar concentration value. The advection-diffusion equations (\ref{advdiff:eq}) are discretized as follows, 

\begin{equation}
\label{Discrete:AdvDiff} c^{k+1} = c^{k} + \Delta t \bigg( D\ \textbf{D}^{2,0} c^{k} - \textbf{D}^0 \textbf{u}^k \cdot \tilde{\textbf{D}}_0^{\pm} c^{k} \bigg), 
\end{equation}

where $D$ is the diffusion coefficient, $\textbf{D}^{2,0}$ is the central differencing operator for second derivatives, and $\tilde{\textbf{D}}_0^{\pm}$ is the upwind differencing operator which depends on the sign of $c$ at that particular point in time. We explicitly define $\textbf{D}^{2,0}$, as follows,

\begin{equation}
\label{Discrete:2ndDiff} \textbf{D}^{2,0} = \left(  D_{1}^{2,0}, D_{2}^{2,0} \right)
\end{equation}
with 
\begin{equation}
\label{Discrete:2ndDiff_2} \left( D_{\alpha}^{2,0}\phi \right)(\textbf{x}) = \frac{ \phi\big( \textbf{x} + \Delta \textbf{x} e_{\alpha} \big) -2 \phi\big(\textbf{x}\big) + \phi\big( \textbf{x} - \Delta\textbf{x} e_{\alpha} \big) }{\Delta x^2}.
\end{equation}

The upwind operator is defined as 
\begin{equation}
\label{Discrete:Upwind_1} \tilde{\textbf{D}}_0^{\pm} = \left( \tilde{D}_1^{\pm}, \tilde{D}_2^{\pm} \right)
\end{equation}
with
\begin{equation}
\label{Discrete:Upwind_2} \left( \tilde{D}_{\alpha}^{\pm}\phi \right)(\textbf{x}) = \left\{  \begin{array}{c} \frac{ \phi\big( \textbf{x} + \Delta \textbf{x} e_{\alpha} \big) - \phi\big( \textbf{x} \big) }{\Delta x}, \ \ \ \ \phi(\textbf{x})\leq 0 \\ \\
\frac{ \phi\big( \textbf{x} \big) - \phi\big( \textbf{x} - \Delta\textbf{x} e_{\alpha} \big) }{\Delta x}, \ \ \ \ \phi(\textbf{x})>0. \end{array}\right.
\end{equation}


%
%


\bibliographystyle{elsarticle-num}

\bibliography{heart}

\begin{thebibliography}{10}
\expandafter\ifx\csname url\endcsname\relax
  \def\url#1{\texttt{#1}}\fi
\expandafter\ifx\csname urlprefix\endcsname\relax\def\urlprefix{URL }\fi
\expandafter\ifx\csname href\endcsname\relax
  \def\href#1#2{#2} \def\path#1{#1}\fi

\bibitem{Bathe:2008}
K.~J. Bathe, Fluid-structure interactions, Mechanical Engineering April (2008)
  67--68.

\bibitem{Peskin:1972}
C.~Peskin, Flow patterns around heart valves: A numerical method, J. Comput.
  Phys. 10(2) (1972) 252--271.

\bibitem{Peskin:1977}
C.~Peskin, Numerical analysis of blood flow in the heart, J. Comput. Phys. 25
  (1977) 220--252.

\bibitem{GriffithThesis:2005}
B.~E. Griffith, Simulating the blood-muscle-vale mechanics of the heart by an
  adaptive and parallel version of the immsersed boundary method (ph.d.
  thesis), Courant Institute of Mathematics, New York University.

\bibitem{Hieber:2008}
S.~Hieber, P.~Koumoutsakos, An immersed boundary method for smoothed particle
  hydrodynamics of self-propelled swimmers, J. Comput. Phys. 227 (2008)
  8636–8654.

\bibitem{Hoover:2015}
A.~P. Hoover, L.~A. Miller, A numerical study of the benefits of driving
  jellyfish bells at their natural frequency, J. Theor. Biol. 374 (2015)
  13--25.

\bibitem{Miller:2004}
L.~A. Miller, C.~S. Peskin, When vortices stick: an aerodynamic transition in
  tiny insect flight, J. Exp. Biol. 207 (2004) 3073–3088.

\bibitem{Miller:2009}
L.~A. Miller, C.~S. Peskin, A computational fluid dynamics of clap and fling in
  the smallest insects, J. Exp. Biol. 208 (2009) 3076--3090.

\bibitem{SJones:2015}
S.~K. Jones, R.~Laurenza, T.~L. Hedrick, B.~E. Griffith, L.~A. Miller, Lift-
  vs. drag-based for vertical force production in the smallest flying insects,
  J. Theor. Biol. 384 (2015) 105--120.

\bibitem{Tytell:2010}
E.~Tytell, C.~Hsu, T.~Williams, A.~Cohen, L.~Fauci, Interactions between
  internal forces, body stiffness, and fluid environment in a neuromechanical
  model of lamprey swimming, Proc. Natl. Acad. Sci. 107 (2010) 19832–19837.

\bibitem{Battista:2015}
N.~A. Battista, A.~J. Baird, L.~A. Miller, A mathematical model and matlab code
  for muscle-fluid-structure simulations, Integr. Comp. Biol.

\bibitem{Hamlet:2015}
C.~Hamlet, L.~J. Fauci, E.~D. Tytell, The effect of intrinsic muscular
  nonlinearities on the energetics of locomotion in a computational model of an
  anguilliform swimmer, J. Theor. Biol. 385 (2015) 119--129.

\bibitem{Zhu:2011}
L.~Zhu, G.~He, S.~Wang, L.~A. Miller, X.~Zhang, Q.~You, S.~Fang, An immersed
  boundary method by the lattice boltzmann approach in three dimensions,
  Computers and Mathematics with Applications 61 (2011) 3506–3518.

\bibitem{Kramer:2008}
P.~R. Kramer, C.~S. Peskin, P.~J. Atzberger, On the foundations of the
  stochastic immersed boundary method, Comp. Meth. in Appl. Mech. and Eng. 197
  (2008) 2232--2249.

\bibitem{Fogelson:2008}
A.~L. Fogelson, R.~D. Guy, Immersed-boundary-type models of intravascular
  platelet aggregation, Comput. Methods Appl. Mech. Engrg. 197 (2008)
  2087–2104.

\bibitem{Lee:2010}
P.~Lee, B.~E. Griffith, C.~S. Peskin, The immersed boundary method for
  advection-electrodiffusion with implicit timestepping and local mesh
  refinement, J. Comp. Phys. 229(13) (2010) 5208--5227.

\bibitem{Strychalski:2012}
W.~Strychalski, R.~D. Guy, Viscoelastic immersed boundary methods for zero
  reynolds number flow, Comm. in Comp. Phys. 12 (2012) 462--478.

\bibitem{Du:2014}
J.~Du, R.~D. Guy, A.~L. Fogelson, An immersed boundary method for two-fluid
  mixtures, J. Comp. Phys. 262 (2014) 231--243.

\bibitem{Baird:2015}
A.~J. Baird, L.~D. Waldrop, L.~A. Miller, Neuromechanical pumping: boundary
  flexibility and traveling depolarization waves drive flow within valveless,
  tubular hearts, Japan J. Indust. Appl. Math. 32 (2015) 829--846.

\bibitem{Waldrop:2015}
L.~D. Waldrop, L.~A. Miller, The role of the pericardium in the valveless,
  tubular heart of the tunicate, \emph{{C}iona savignyi}, J. Exp. Biol. 218
  (2015) 2753--2763.

\bibitem{Fogelson:IBIS}
D.~J. Eyre, A.~L. Fogelson, \href{http://www.math.utah.edu/IBIS/}{Ibis: A
  software system for immersed boundary and interface simulations} (1997).
\newline\urlprefix\url{http://www.math.utah.edu/IBIS/}

\bibitem{BGriffithIBAMR}
B.~E. Griffith, \href{https://github.com/IBAMR/IBAMR}{An adaptive and
  distributed-memory parallel implementation of the immersed boundary (ib)
  method} (2014) [cited October 21, 2014].
\newline\urlprefix\url{https://github.com/IBAMR/IBAMR}

\bibitem{Griffith:IBFE}
B.~E. Griffith, X.~Luo, Hybrid finite difference/finite element version of the
  immersed boundary method, Int. J. Numer. Meth. Engng. 0 (2012) 1--26.

\bibitem{PETSc}
H.~Zhang, \href{http://www.mcs.anl.gov/petsc}{{PETSc}: Portable, extensible
  toolkit for scientific computation} (2009).
\newline\urlprefix\url{http://www.mcs.anl.gov/petsc}

\bibitem{SAMRAI}
{Lawrence Livermore National Laboratory},
  \href{http://www.llnl.gov/CASC/SAMRAI}{{SAMRAI}: Structured adaptive mesh
  refinement application infrastructure} (2007).
\newline\urlprefix\url{http://www.llnl.gov/CASC/SAMRAI}

\bibitem{libMesh}
B.~S. Kirk, J.~W. Peterson, R.~H. Stogner, G.~F. Carey, {\texttt{libMesh}: A
  C++ Library for Parallel Adaptive Mesh Refinement/Coarsening Simulations},
  Engineering with Computers 22~(3-4) (2006) 237--254,
  \url{http://dx.doi.org/10.1007/s00366-006-0049-3}.

\bibitem{openMPI}
E.~Gabriel, G.~E. Fagg, G.~Bosilca, T.~Angskun, J.~J. Dongarra, J.~M. Squyres,
  V.~Sahay, P.~Kambadur, B.~Barrett, A.~Lumsdaine, R.~H. Castain, D.~J. Daniel,
  R.~L. Graham, T.~S. Woodall, Open {MPI}: Goals, concept, and design of a next
  generation {MPI} implementation, in: Proceedings, 11th European PVM/MPI
  Users' Group Meeting, Budapest, Hungary, 2004, pp. 97--104.

\bibitem{Wiens:MATIB}
B.~D. Froese, J.~Wiens, \href{https://github.com/eldila/MatIB}{Matib: A simple
  immersed boundary method solver in matlab.} (2013).
\newline\urlprefix\url{https://github.com/eldila/MatIB}

\bibitem{Mesnard:pyIBM}
O.~Mesnard, \href{https://github.com/mesnardo/pyIBM}{pyibm: An immersed
  boundary method python code} (2014).
\newline\urlprefix\url{https://github.com/mesnardo/pyIBM}

\bibitem{Peskin:IB}
C.~S. Peskin,
  \href{https://www.math.nyu.edu/faculty/peskin/ib_lecture_notes/index.html}{The
  immersed boundary method} (2008).
\newline\urlprefix\url{https://www.math.nyu.edu/faculty/peskin/ib_lecture_notes/index.html}

\bibitem{Peskin:2002}
C.~S. Peskin, The immersed boundary method, Acta Numerica 11 (2002) 479--517.

\bibitem{MATLAB:2015a}
MATLAB, version 8.5.0 (R2015a), The MathWorks Inc., Natick, Massachusetts, USA,
  2015.

\bibitem{Python:Python}
G.~{Van Rossum}, Python, version 3.5, https://www.python.org, 2015.

\bibitem{Chorin:1967}
A.~J. Chorin, The numerical solution of the navier-stokes equations for an
  incompressible fluid, Bull. Am. Math. Soc. 73 (1967) 928--931.

\bibitem{Brown:2001}
D.~L. Brown, R.~Cortez, M.~L. Minion, Accurate projection methods for the
  incompressible navier–stokes equations, J. Comp. Phys. 168 (2001) 464--499.

\bibitem{Cooley:1965}
J.~W. Cooley, J.~W. Tukey, An algorithm for the machine calculation of complex
  fourier series, Math. Comput. 19 (1965) 297--301.

\bibitem{Press:1992}
W.~H. Press, B.~P. Flannery, S.~A. Teukolsky, W.~T. Vetterling, Fast fourier
  transform, Ch. 12 in Numerical Recipes in FORTRAN: The Art of Scientific
  Computing 2 (1992) 490--529.

\bibitem{Hamlet:2012}
C.~Hamlet, L.~A. Miller, Feeding currents of the upside-down jellyfish in the
  presence of background flow, Bull. Math. Bio. 74(11) (2012) 2547--2569.

\bibitem{Stockie:2009}
J.~M. Stockie, Modelling and simulation of porous immersed boundaries,
  Computers and Structures 87 (2009) 701--709.

\bibitem{Peskin:1993}
C.~S. Peskin, B.~F. Printz, Improved volume conservation in the computation of
  flows with immersed elastic boundaries, J. Comput. Phys. 105 (1993) 33--46.

\bibitem{Kim:2006}
Y.~Kim, C.~S. Peskin, 2d parachute simulation by the immersed boundary method,
  SIAM J. Sci. Comput. 28 (2006) 2294–2312.

\bibitem{Hill:1938}
A.~V. Hill, The heat of shortening and the dynamic constants of muscle, Proc.
  R. Soc. Lond. 126 (1938) 136--195.

\bibitem{Fung:1993}
Y.~C. Fung, Biomechanics: mechanical properties of living tissues,
  Springer-Verlag, New York, USA, 1993.

\bibitem{Hatze:1981}
H.~Hatze, A comprehensive model for human motion simulation and its application
  to the take-off phase of the long jump, J. Biomech. 14 (1981) 135--142.

\bibitem{Challis:1994}
J.~H. Challis, D.~G. Kerwin, Determining individual muscle forces during
  maximal activity: Model development, parameter determination, and validation,
  Hum. Movement Sci. 13 (1994) 29–61.

\bibitem{Santhanakrishnan:2009}
A.~Santhanakrishnan, N.~Nguyen, J.~Cox, L.~A. Miller, Flow within models of the
  vertebrate embryonic heart, J. Theor. Biol. 259 (2009) 449--461.

\bibitem{Newren:2007}
E.~P. Newren, A.~L. Fogelson, R.~D. Guy, R.~M. Kirby, Unconditionally stable
  discretizations of the immersed boundary equations, J. Comput. Phys. 222
  (2007) 702--719.

\bibitem{Teran:2009}
J.~M. Teran, C.~S. Peskin, Tether force constraints in stokes flow by the
  immersed boundary method on a periodic domain, SIAM J. Sci. Comput. 31(5)
  (2009) 3404--3416.

\bibitem{Teran:2010}
J.~M. Teran, L.~Fauci, M.~Shelley, Viscoelastic fluid response can increase the
  speed and efficiency of a free swimmer, Phys. Rev. Lett. 104(3) (2010)
  038101.

\bibitem{Griffith:2013}
B.~E. Griffith, C.~S. Peskin, Electrophysiology, Comm. Pure Appl. Math. 66
  (2013) 1837--1913.

\bibitem{Williams:1998}
T.~L. Williams, G.~Bowtell, N.~A. Curtin, Predicting force generation by
  lamprey muscle during applied sinusoidal movement using a simple dynamic
  model, J Exp Biol. 201 (1998) 869--875.

\bibitem{McMillen:2008}
T.~McMillen, T.~Williams, P.~Holmes, Nonlinear muscles, passive viscoelasticity
  and body taper conspire to create neuromechanical phase lags in anguilliform
  swimmers, PLoS Comp. Bio. 4(8) (2008) e1000157.

\bibitem{Schroder:2006}
W.~Schroeder, K.~Martin, B.~Lorensen, The Visualization Toolkit (4th edition),
  Kitware, Carrboro, USA, 1993.

\bibitem{Paraview:2005}
J.~Ahrens, B.~Gerveci, C.~Law, ParaView: An End-User Tool for Large Data
  Visualizations, Elsevier, Atlanta, USA, 2005.

\bibitem{HPV:VisIt}
H.~Childs, E.~Brugger, B.~Whitlock, J.~Meredith, S.~Ahern, D.~Pugmire,
  K.~Biagas, M.~Miller, C.~Harrison, G.~H. Weber, H.~Krishnan, T.~Fogal,
  A.~Sanderson, C.~Garth, E.~W. Bethel, D.~Camp, O.~R\"{u}bel, M.~Durant, J.~M.
  Favre, P.~Navr\'{a}til, {VisIt: An End-User Tool For Visualizing and
  Analyzing Very Large Data}, in: {High Performance Visualization--Enabling
  Extreme-Scale Scientific Insight}, 2012, pp. 357--372.

\bibitem{Kiger:2016}
K.~Kiger, J.~Westerweel, C.~Poelma, {Introduction to Particle Image
  Velocimetry},
  \url{http://www2.cscamm.umd.edu/programs/trb10/presentations/PIV.pdf} (2016).

\bibitem{Dickinson:1993}
M.~H.~D. K, Gotz, Unsteady aerodynamic performance of model wings at low
  reynolds numbers, J. Exp. Biol. 174 (1993) 45--64.

\bibitem{Birch:2004}
J.~Birch, W.~Dickson, M.~H. Dickinson, Force production and flow structure of
  the leading edge vortex at high and low reynolds numbers, J. Exp. Biol. 207
  (2004) 1063--1072.

\bibitem{Crowl:2011}
L.~M. Crowl, A.~L. Fogelson, Analysis of mechanisms for platelet near-wall
  excess under arterial blood flow conditions, J. Fluid Mech. 676 (2011)
  348--375.

\bibitem{Zhang:2014}
C.~Zhang, R.~D. Guy, B.~Mulloney, Q.~Zhang, T.~J. Lewis, The neural mechanism
  of optimal limb coordination in crustacean swimming, PNAS 111 (2014)
  13840--13845.

\bibitem{Lewis:2015}
O.~L. Lewis, S.~Zhang, R.~D. Guy, J.~{del Alamo}, Coordination of
  contractility, adhesion and flow in migrating physarum amoebae, J. R. Soc.
  Interface 12(106).

\bibitem{openFOAM:openFOAM}
O.~Foundation, \href{https://github.com/OpenFOAM}{Official openfoam repository}
  (2014).
\newline\urlprefix\url{https://github.com/OpenFOAM}

\bibitem{COMSOL:2011}
C.~Multiphysics, version 4.2 (R2015a), COMSOL Inc., Burlington, MA, USA, 2015.

\bibitem{Fluent:2016}
C.~ANSYS~Fluent, CFD, version 17.0, ANSYS Inc., Cecil Township, Pennslyvania,
  USA, 2016.

\bibitem{Lee:2003}
L.~Lee, R.~J. Leveque, An immersed interface method for incompressible
  navier-stokes equations, SIAM J. SCI. COMPUT. 25(3) (2003) 832–856.

\bibitem{Li:2003}
Z.~Li, An overview of the immersed interface method and its applications,
  Taiwanese J. Math. 7(1) (2003) 1--49.

\bibitem{Ubbink:1999}
O.~Ubbink, R.~I. Issa, Method for capturing sharp fluid interfaces on arbitrary
  meshes, J. Comp. Phys. 153 (1999) 26--50.

\bibitem{Udaykumar:2001}
H.~Udaykumar, R.~Mittal, P.~Rampunggoon, A.~Khanna, A sharp interface cartesian
  grid method for simulating flows with complex moving boundaries, J. Comp.
  Phys. 20 (2001) 345--380.

\bibitem{Cortez:2000}
R.~Cortez, M.~Minion, The blob projection method for immersed boundary
  problems, J. Comp. Phys. 161 (2000) 428–453.

\bibitem{Sethian:1999}
J.~A. Sethian, Level Set Methods and Fast Marching Methods : Evolving
  Interfaces in Computational Geometry, Fluid Mechanics, Computer Vision, and
  Materials Science, Cambridge University Press, Cambridge, UK, 1999.

\bibitem{Osher:2002}
S.~J. Osher, R.~Fedkiw, Level Set Methods and Dynamic Implicit Surfaces,
  Springer-Verlag, New York, NY, USA, 2002.

\bibitem{Peskin:1996}
C.~S. Peskin, D.~M. McQueen, Fluid dynamics of the heart and its valves, in:
  F.~R. Adler, M.~A. Lewis, J.~C. Dalton (Eds.), Case Studies in Mathematical
  Modeling: Ecology, Physiology, and Cell Biology, Prentice-Hall, New Jersey,
  1996, Ch.~14, pp. 309--338.

\end{thebibliography}

\end{document}